\documentclass[aps,pra,reprint,groupedaddress]{revtex4-1}
\usepackage{graphicx}
\usepackage{color}
\graphicspath{{figure/}}
\usepackage{amsmath}
\usepackage{hyperref}
\graphicspath{{figure/}}
\begin{document}

\title{Noise budget of a trapped on-chip cold-atom rubidium 87 clock}

\author{M. Dupont-Nivet$^{1}$\footnote{Corresponding author: matthieu.dupontnivet@thalesgroup.com}, B. Wirtschafter$^{1}$, S. Hello$^{2,3}$, C. I. Westbrook$^{3}$}

\affiliation{${}^{1}$Thales Research and Technology France, 1 av. Augustin Fresnel, 91767 Palaiseau, France \\
${}^{2}$Thales AVS France SAS, 40 rue de la Brelandière, 86100 Châtellerault, France \\
${}^{3}$Laboratoire Charles Fabry, Institut d'Optique Graduate School, 2 av. Augustin Fresnel, 91127 Palaiseau France }

\date{\today}

\begin{abstract}
In this paper, we present a realization of an on chip atomic clock using a cold cloud of rubidium 87 atoms. 
This clock is based on a Ramsey interferometer with a Ramsey time around 600~ms. 
This is realized with large laboratory temperature drift during the measurement (few degrees per day) and without magnetic-field shielding. 
We review the experimental implementation of this clock and give a full study of the known noises present in this atomic clock.
\end{abstract}
\pacs{}

\maketitle

%%%%%%%%%%%%%%%%%%%%
%    Introduction
%%%%%%%%%%%%%%%%%%%%

\section{Introduction}

Cold atom technology has interesting possibilities to realize clocks \cite{Abgrall2015,Koller2017,Ludlow2015}, accelerometers \cite{Pelle2013,Alauze2018,Xu2019,Gillot2014,Hu2013,Kasevich2013,Chen2019,Perrin2019}  and gyroscopes \cite{Dutta2016,Canuel2006,Durfee2006,Savoie2018} with high accuracy, stability and sensitivity.
Despite their complexity, atom chips \cite{Reichel2010,Reichel1999,Reichel2002} hold great promise for compact accelerometers \cite{Bohi2009,Ammar2014,DupontNivet2014} and gyroscopes, \cite{GarridoAlzar2012,Horne2017} in particular for rubidium atoms. 
They have already been used to realize atomic clocks \cite{Treutlein2004,Szmuk2015}, and protocols have been presented to measure acceleration \cite{Ammar2014,DupontNivet2014} and rotation rates \cite{Wirtschafter2022,DupontNivet2015b,Wirtschafter2022,DupontNivet2015b,Navez2016,Pandey2019}. 
Our implementation of an accelerometer and gyroscope uses a thermal cold cloud of rubidium 87 atoms that are trapped in a harmonic magnetic trap created with DC currents running through wires on an atom chip. 
Two magnetically trappable states of rubidium 87, namely, $\left|a\right>=\left|F=1,m_F=-1\right>$ and $\left|b\right>=\left|F=2,m_F=1\right>$, are used in a Ramsey interferometer that takes place in this magnetic trap. 
These two states are selected because they can be trapped using DC magnetic fields keeping them trapped during the whole Ramsey sequence. 
Near a magnetic field of 3.229~G (called the magic field), the transition frequency between them is less sensitive to magnetic field \cite{Harber2002,Treutlein2004}, for example for magnetic fluctuations of 1~mG the transition frequency fluctuations are reduced by about three order of magnitudes. 
This leads to the realization of on-chip Ramsey clocks.
Between the two $\pi/2$ pulses of the Ramsey protocol, let us add a spatial motion of $\left|a\right>$ and $\left|b\right>$ to split, held apart and merge them using two microwave near fields which create two state selective potentials, one for each state \cite{Ammar2014}. 
If this displacement of the two states is done in opposite direction along a straight line, the interferometer becomes sensitive to accelerations along that direction \cite{Ammar2014,DupontNivet2014}.
In addition, if the minimum of the harmonic magnetic trap is moved in a direction noncollinear with the previous straight line, the two states can be displaced in opposite directions along a path that encloses a nonzero area, thus the interferometer becomes sensitive to rotation rates \cite{DupontNivet2015b}.  
Therefore, accelerometers and gyroscopes based on the previously described implementation are on chip cold atom clocks where we added a spatial motion of the atoms, thus they have a significant overlap of their noise budget with the on-chip atomic clock one.

While previous realizations of on-chip atomic clocks favorably compare with ours in terms of relative frequency stablity 
-- reference \cite{Szmuk2015} reported a $2\cdot 10^{-13}$ relative stability at one shot and we report only a $1.4\cdot 10^{-12}$ relative stability -- 
the focus of this paper is not to report the best in class on-chip atomic clock, but to report the on-chip atomic clock noise budget, as a preliminary step of the noise budget of an on-chip accelerometer or gyroscope.
Three main reasons explain this difference of relative frequency stability: (i)~reference \cite{Szmuk2015} used a local oscillator giving a Dick effect one order of magnitude lower than ours, (ii)~\cite{Szmuk2015} used a magnetic shielding and ultrastable current sources while we are using off-the-shelf current sources and no magnetic shielding leading to a magnetic-field noise around 0.6~mG while \cite{Szmuk2015} achieved a level of 16~$\mu$G and (iii)~Identical Spin Rotation Effect (ISRE) \cite{Deutsch2010} in reference \cite{Szmuk2015} led to a Ramsey time of 5~s while we are using a Ramsey time around 600~ms.

The goal of this paper is twofold: showing the results obtained with our on chip cold atom clock apparatus and giving a comprehensive review of all the noise identified in this clock implementation as well as possibilities to reduce them. 

The paper is organized as follows, section \ref{sec_experiment} describes the experimental set-up and protocol for measuring the Allan standard deviation and shows the results. 
Section \ref{sec_noise} shows the optimization of the clock stability versus the magnetic field at the bottom of the trap. 
Section \ref{sec_NoiseBudget} gives theoretical description and measurements of all the known sources of noise of this clock. 
Section \ref{sec_discussion} discusses previous results and gives insight of how to reduce the noise. 

%%%%%%%%%%%%%%%%%%%%
%    Experimental protocol
%%%%%%%%%%%%%%%%%%%%

\section{Experimental protocol}
\label{sec_experiment}

\begin{figure}
\centering  \includegraphics[width=0.48\textwidth]{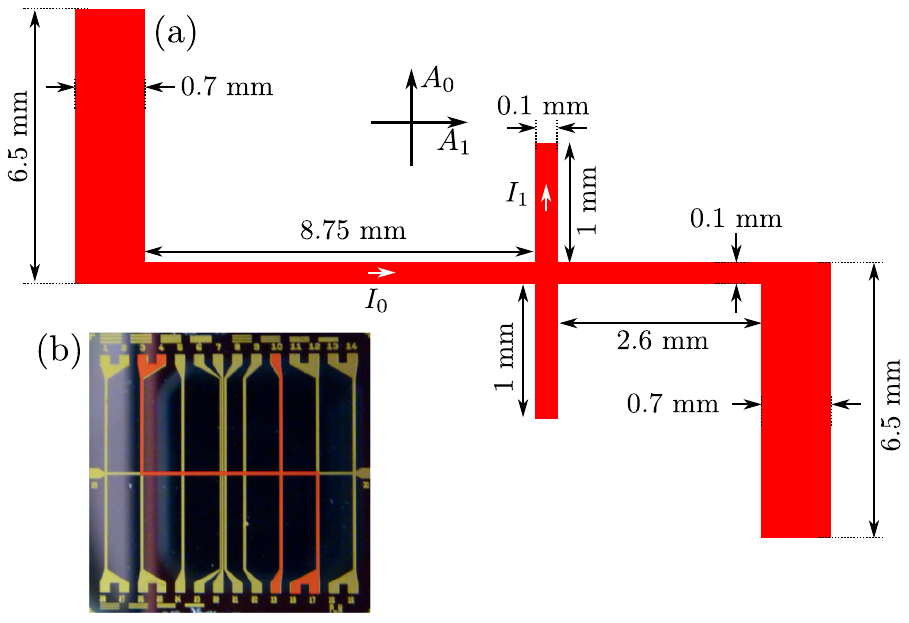}
\caption{\label{fig_AtomChip} (Color online) (a) Layout of the atom chip wires used for the dimple trap during the Ramsey sequence on the chip. 
(b) Photograph of the atom chip. 
The wires highlighted in red are those used for the dimple trap of panel (a).
}
\end{figure}

\subsection{Atom preparation}

We use the apparatus described in \cite{Huet2012,Huet2013,DupontNivet2016}. 
A cloud of around $5\cdot10^8$ rubidium 87 atoms is first cooled down in a three-dimensional magneto-optical trap using a beam configuration as described in \cite{Farkas2010,Squires2008} to accommodate the presence of an atom chip \cite{Reichel2010}. In this configuration, the three pairs of retroreflected beams are not orthogonal. 
This three-dimensional magneto-optical trap is loaded in less than 500~ms using a cold atom beam created with a two-dimensional magneto-optical trap \cite{Dieckmann1998,Schoser2002}. 
The atoms are further cooled to 40~$\mu$K by polarization gradient cooling and optically pumped into state $\left|F=2,m_F=2\right>$. 
After optical pumping the atoms are loaded into a dimple-shaped magnetic trap created by the atom chip, the loading protocol is described in \cite{DupontNivet2016,Squires2008}. 
The last cooling stage is a 1.5~s radio-frequency evaporation until the cloud reaches a temperature few times the Bose-Einstein condensation threshold. 
The atomic cloud is transferred into internal state $\left|b\right>$ using microwave  stimulated Raman adiabatic passage (STIRAP) \cite{DupontNivet2015,Vitanov2017}. 
After the STIRAP there are on average $N_i=(39\pm 2)\cdot 10^3$ atoms at a temperature of $T=500$~nK available for the Ramsey interferometer.

\subsection{Detection}

\begin{figure}
\centering  \includegraphics[width=0.40\textwidth]{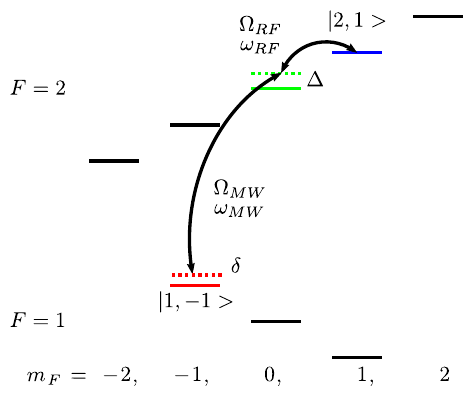}
\caption{(Color online) Levels of rubidium 87 ground state. 
The two levels $\left|a\right>=\left|F=1,m_F=-1\right>$ (in red) and $\left|b\right>=\left|F=2,m_F=1\right>$ (in blue) are used as the two clock levels for the Ramsey interferometer. 
Two frequencies drive the transition: a microwave near $f_{MW}=\omega_{MW}/(2\pi)\approx$~6.8~GHz with Rabi frequency $\Omega_{MW}$ and a radio frequency near $f_{RF}=\omega_{RF}/(2\pi)\approx$~1~MHz with Rabi frequency $\Omega_{RF}$.
$\Delta$ and $\delta$ are the one and two photon detunings respectively.}
\label{fig_RbLevel}
\end{figure}

Atoms in both clock states, $\left|b\right>$ and $\left|a\right>$ are detected by absorption imaging after the trap is turned off and a short magnetic-field gradient is pulsed on. The gradient moves them away from the chip surface more rapidly that the effect of gravity alone. 
A first detection pulse (35~$\mu$s) comes 5~ms after releasing the atoms. 
It is tuned to the $F=2$ and $F'=3$ levels and thus interacts with the atoms in the $\left|b\right>$ state. 
Next, a 200~$\mu$s repumping pulse transfers the atoms in the $\left|a\right>$ state to the $F=2$ level and after an additional 1.5~ms, the detection beam is again pulsed on to record the absorption of the transferred atoms. 
Because of the different times of flight, the two absorption profiles can be recorded on the same camera image. 

The time between the two detection pulses is chosen such that \cite{DupontNivet2016,Wirtschafter2022}: (i)~atoms in the $\left|a\right>$ state have fallen sufficiently far so as not to overlap the atoms in the $\left|b\right>$ state during their detection and 
(ii)~atoms in the $\left|b\right>$ state have moved out of the depth of field of the imaging lens due to the acceleration by the first detection pulse. 
This leads to a first image called $I_A$ with two separate absorption profiles, one for each clock state. 
A second image called $I_W$ is taken 4~ms after $I_A$ in exactly the same conditions but without the atoms. 
The two images allow us to compute the optical density profiles of each state \cite{Ketterle1999,Lewandowski2003,DupontNivet2016}. 

\subsection{Reduction of the noise of the optical density images}
\label{sec_ODNoiseReduc}

Although the delay between the absorption and reference images is very short (4~ms), the interference patterns in the images have had enough time to change significantly leading to unwanted artifacts in the optical density image. 
In the following we describe our protocol to reduce these artifacts, inspired by the eigen face method \cite{Turk1991,Turk1991b,Pissarenko2002,Sirovich1987}. 
This procedure allows us to reliably deduce the fraction of atoms in the two clock states. 

Suppose that we have already collected $N_I$ reference images $\left(I_W^k\right)_{k\in [1,N_I]}$ over the previous experimental acquisitions. 
In the areas of an image $I_A$ where there are no atoms, $I_A$ should be equal to $I_W$. 
Thus to compute the optical density we construct a corrected reference image $I_W^{cor}$ with the previously collected $\left(I_W^k\right)_{k\in [1,N_I]}$ images \cite{Li2007b,Deutsch2011,Ockeloen2010,Ockeloen2010b,Ockeloen2014}:
\begin{equation}
I_W^{cor} = \sum_{k=1}^{N_I} c_k I_W^k \,,
\end{equation}
where the weights $c_k$ are real and determined from the minimization of the following cost function:
\begin{equation}
J=\sum_{(i,j)} \left[ M(i,j)\ast I_A(i,j) - \sum_{k=1}^{N_I} c_k M(i,j)\ast I_W^k(i,j) \right]^2 \,,
\end{equation}
where $\ast$ is the term-by-term matrix product, $(i,j)$ runs over all the pixel of the images and $M$ is a mask image which is equal to one in the area where we know that there are no atoms and zero otherwise. 
In practice, our images are 226 by 101 pixels, we use about $N_I=700$ images $I_W^k$ and $M$ contains around 12~000 pixels equal to one. 
Then $I_A$ and $I^{cor}_W$ are used to compute the optical density, and we typically reduce the amplitude of the artifacts in the optical density by a factor between 5 and 10.

The panel of reference images $\left(I_W^k\right)_{k\in [1,N_I]}$ is updated after each experimental run by discarding the oldest image of the panel and adding to it the latest acquired reference image $I_W$.  

\subsection{Ramsey interferometer and clock frequency}
\label{sec_RamseyInter}

\subsubsection{Choosing the Rabi pulse parameters}

\begin{figure*}
\centering  \includegraphics[width=1.00\textwidth]{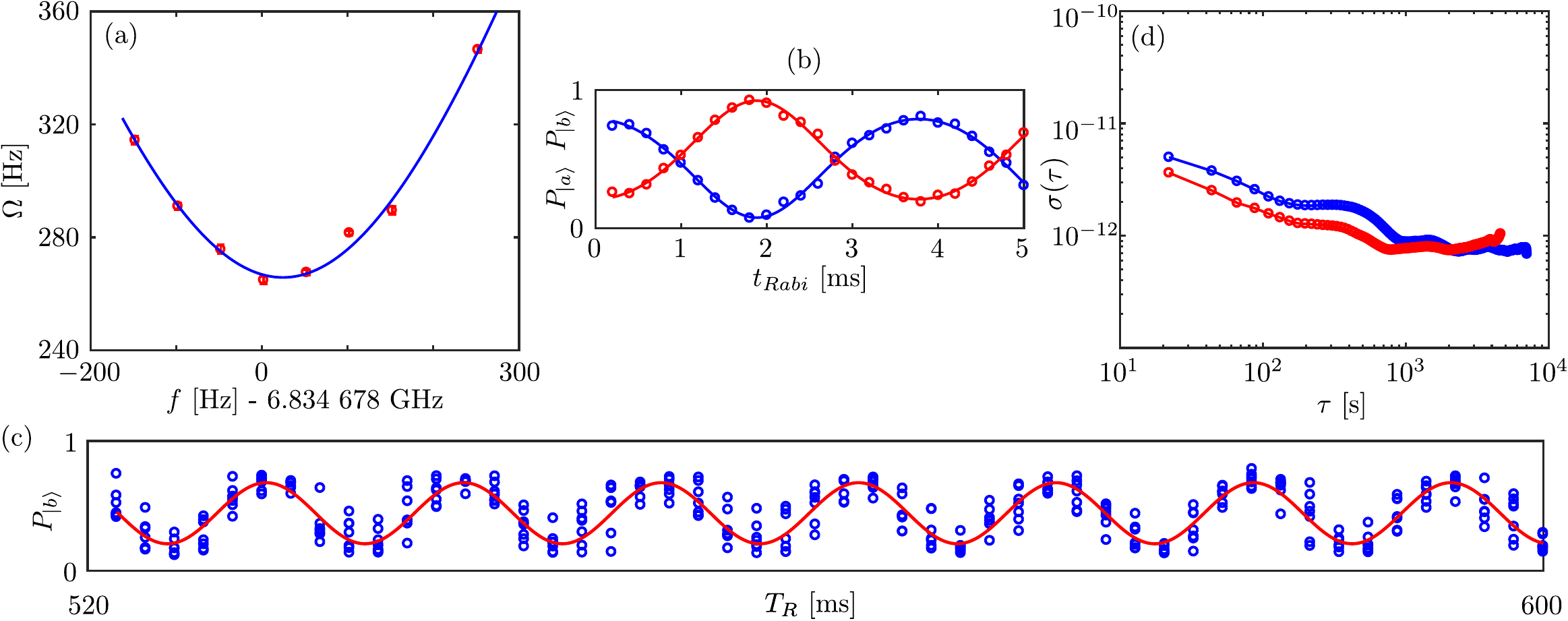}
\caption{\label{fig_Allan_Result_1} (Color online) Main experimental steps leading to the Allan deviation (d) at $B_0=3.160$~G. 
(i) In panel (a) for the frequency of the two photon field driving the Rabi oscillations recorded in panel (b), (ii) in panel (b)determination of the Rabi pulses duration for panel (c), and (iii) in panel (c) record of the Ramsey fringes to chose the Ramsey time for measuring Allan deviation in panel (d).
(a) Generalized Rabi frequency $\Omega$~[Hz] as the function of the two photons driving field frequency $f=f_{RF}+f_{MW}$~[Hz]. 
Red open circles are experimental data and the solid blue line is a parabolic fit (see text). 
(b) Rabi oscillation of populations $P_{\left|b\right>}$ in blue and $P_{\left|a\right>}$ in red as a function of the Rabi time $t_{Rabi}$~[ms]. 
Open circles are experimental data and the solid lines are sinusoidal fit. 
(c) Ramsey fringes: population $P_{\left|b\right>}$ as a function of the Ramsey time $T_R$~[ms]. 
Open blue circles are experimental data and the solid red line is a sinusoidal fit (see text).
(d) Allan deviation $\sigma(\tau)$ as a function of the integration time $\tau$~[s]. 
Data in blue were taken on the 22$^{nd}$ of November 2016 and in red on the 15$^{th}$ of November 2016. 
The open circles are experimental data and the solid lines are a guide for the eyes. }
\end{figure*}

The clock interrogation takes place in a dimple magnetic trap with the following parameters (see figure \ref{fig_AtomChip} for the notational definitions): $I_0=278$~mA, $I_1=102$~mA, $A_0\approx 4.5$~G and $A_1\approx 5.3$~G. 
This results in a trap with a $B_0\approx3.160\pm 0.005$~G magnetic field at the bottom of the trap which is slightly shifted from the magic field (3.229~G) \cite{Rosenbusch2009}, trap eigenfrequencies of $(\omega_x,\omega_y,\omega_z)\approx2\pi\times(85,150,160)$~Hz (frequencies are given for atoms in state $\left|b\right>$) and a trap to surface distance of approximately 100~$\mu$m. 
As shown in figure \ref{fig_RbLevel}, the $\pi/2$ pulses of the Ramsey interferometer are created using two photon Rabi oscillations \cite{Gentile1989}. 
The frequencies of the two photons are adjusted for having a one photon detuning $\Delta=1.209$~MHz and a zero two photon detuning when the Rabi pulse is on, leading to $f_{MW}=6.833~678~024$~GHz and $f_{RF}=1$~MHz. 
For this optimisation we first fixed $f_{RF}$ at 1~MHz and for different values of $f_{MW}$ we measured the frequency of the Rabi oscillations (see figure \ref{fig_Allan_Result_1}.a). 
This data are adjusted with a parabolic function according to the formula for the two photon Rabi oscillation frequency $\Omega$ given by \cite{Gentile1989}:
\begin{equation}
\Omega = \sqrt{\Omega_R^2+\left(\delta - \delta_{21}\right)^2}
\label{eq_GeneRabiFreq}
\end{equation}
with $\Omega_R=\Omega_{MW}\Omega_{RF}/(2|\Delta|)$ the Rabi frequency of the two photon field driving the transition, $\delta_{21}=\Omega_{MW}^2/(4\Delta)$ the light shift induced by the two photon field and $\delta=\omega-\omega_0$, where $\omega=\omega_{MW}+\omega_{RF}$ is the frequency of the two photon driving field and $\omega_0$ is the atomic transition frequency in the absence of the two photon driving field. When operating at the previously given numerical parameters, which correspond to the minimum of the parabola of figure \ref{fig_Allan_Result_1}.a, we obtain the Rabi oscillations shown in figure \ref{fig_Allan_Result_1}.b. 
This leads to a time of 1~ms for the $\pi/2$ pulses of the Ramsey interferometer.

In figure \ref{fig_Allan_Result_1}.b, the oscillation for state $\left|a\right>$ does not start from $P_{\left|a\right>}=0$ because of offsets on the optical density while detecting atoms in state $\left|a\right>$.
Unfortunately, timing and magnetic-field gradient chosen for the detection lead to two biaises in the atom count in states $\left|a\right>$ and $\left|b\right>$: 
(i)~the STIRAP process is not 100\% efficient so that some atoms remain in state $\left|F=2,m_F=2\right>$ at the beginning of the interrogation. 
Due to the selected imaging timing, they are detected at the same position on the optical density images as state $\left|a\right>$. 
This leads to an offset on the estimation of atom number in state $\left|a\right>$, (ii)~while detecting state $\left|a\right>$ even though atoms in state $\left|b\right>$ have moved out of the imaging lens depth of field and their temperature was dramatically increased by the first detection pulse they still lead to a very small offset in the optical density leading also to an offset on the estimate of the number of atoms in state $\left|a\right>$.

\subsubsection{Recording Ramsey fringes}

Figure \ref{fig_Allan_Result_1}.c shows the population in state $\left|b\right>$ after the interferometry sequence for long interrogation times and for the same excitation frequency as in figure \ref{fig_Allan_Result_1}.b.
The fitting function reads (before starting the interferometer, atoms are prepared in state $\left|b\right>$):
\begin{equation}
P_{\left|b\right>}=\frac{1}{2}\left\{ 1 - \cos\left[\left( \omega - \omega_0 \right) T_R \right] \right\} \,.
\label{eq_PopRamsey}
\end{equation} 
Although the fringes oscillate at $\delta_{21}$ due to our choice of $\omega_{MW}$ and $\omega_{RF}$, the light shift itself does not add noise to the clock frequency (unless it is induced by noisy $\pi /2$ pulses), only noise in the phase $(\omega-\omega_0)T_R$ contributes to the clock signal fluctuations. 
The fit to equation~(\ref{eq_PopRamsey}) gives $\delta_{21}=92.15\pm 0.2$~Hz. 
The minimum of the parabola of figure \ref{fig_Allan_Result_1}.a is found at $f=\omega/(2\pi)=(\omega_{MW}+\omega_{RF})/(2\pi)=6~834~678~024.5\pm 3$~Hz leading to an estimation of $\omega_0/(2\pi) = 6~834~678~116.6\pm 3$~Hz which is in good agreement with the theoretical value of $6~834~678~115.6$~Hz for a magnetic field of $3.160$~G (calculated using the Breit-Rabi formula \cite{Steck2003}). 
Scanning the fringes allows us to chose a Ramsey time $T_R=603.3$~ms so as to operate the interferometer at mid-fringe. 
We then repeat the fringe measurement with the same Ramsey time $T_R$ to build up the Allan standard deviation (figure \ref{fig_Allan_Result_1}.d).
These measurements will be discussed  further in section \ref{sec_AllanVar}.

The time from the beginning of the three-dimensional magneto-optical trap loading to the readout of the interferometer is about 3.5~s, but the measurement is repeated every $\tau_0=22$~s. 
This very conservative total cycle time is chosen to avoid over heating the atom chip, which is only cooled by air convection from the outside.

\subsubsection{Clock frequency as a function of the magnetic field and atom density}

\begin{figure}
\centering  \includegraphics[width=0.48\textwidth]{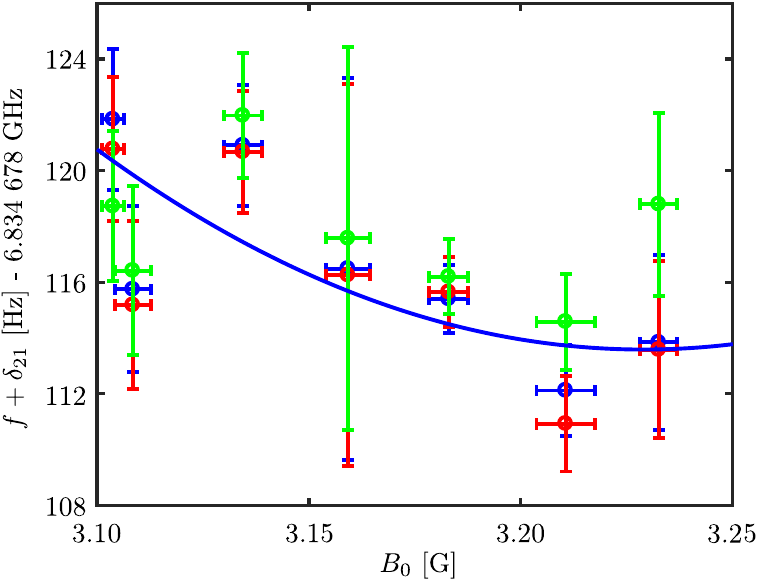}
\caption{\label{fig_Freq_Trans_Coll} (Color online) Measurement of the drift of the clock transition frequency during the Ramsey time induced by collision frequency shift. 
Frequency of the atomic transition [Hz] computed from the value of $f=f_{RF}+f_{MW}$ and $\delta_{21}$ (see section \ref{sec_RamseyInter}) as a function of the magnetic field $B_0$~[G].
Blue open circles are the frequencies fitted on the Ramsey fringes between 0~ms and 80~ms (see figure \ref{fig_RamseyFringes}), red open circles between 240~ms and 320~ms, green open circles between 520~ms and 600~ms. 
The solid blue line is the theoretical value computed from Breit-Rabi formula, without taking into account the density shift.
}
\end{figure} 

Using the Ramsey fringes recorded with the previously described protocol, and as shown in appendix figures \ref{fig_RamseyFringes}.b to \ref{fig_RamseyFringes}.h for several values of the magnetic field at the bottom of the trap, we computed the clock frequency as a function of this magnetic field in figure \ref{fig_Freq_Trans_Coll}. 
Error bars on the frequency are dominated by the error on the determination of the frequency $f_{MW}+f_{RF}$ that minimizes the frequency of the Rabi oscillations (see section \ref{sec_RamseyInter}). 
Nevertheless, within the error bars, the measured frequencies and the computed value from Breit-Rabi formula \cite{Steck2003} agree.
As atoms progressively leave the trap, the density decreases as does the collisional frequency shift, this leads to a change of the frequency of the Ramsey fringes with the Ramsey time. 
In figure \ref{fig_Freq_Trans_Coll}, we display the changes of the clock frequency with the Ramsey time. 
As expected from theory (equation (\ref{eq_ColShiftAverage})) we observe an increase of clock frequency with a decrease of the atom number (except for the lowest magnetic field of the figure \ref{fig_Freq_Trans_Coll}). 

\subsection{Allan variance}
\label{sec_AllanVar}

\begin{figure}
\centering  \includegraphics[width=0.48\textwidth]{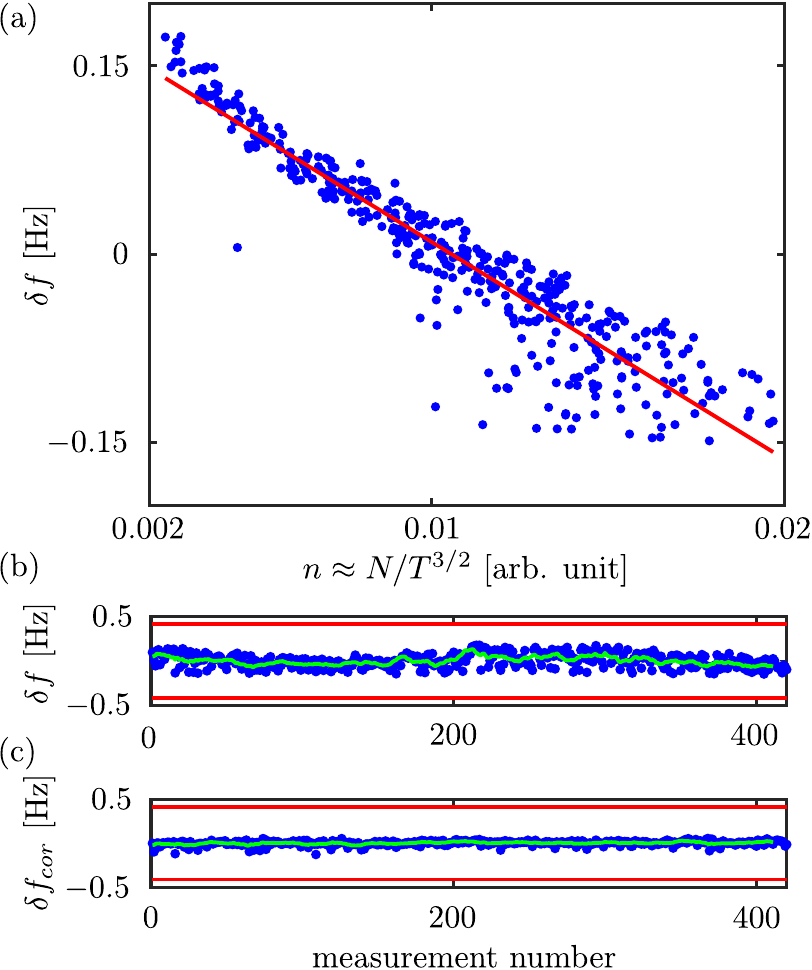}
\caption{\label{fig_cor_density} (Color online) Correlation between the frequency displacement and the atomic density and correction of the measured frequency displacement for the atomic density fluctuations. 
(a) Measured frequency displacement $\delta f$~[Hz] with a Ramsey sequence as a function of the atomic cloud density $n\approx N/T^{3/2}$~[arb. unit]. 
The blue dots are the experimental data, and the solid red line is a linear fit (see text). 
(b) and (c) Measured frequency displacement $\delta f$~[Hz] with a Ramsey sequence as the function of the measurement number. 
The blue dots are the experimental data, the solid green lines are a moving average over ten consecutive experimental data to guide the eyes and the solid red lines show the maximum frequency excursion over half a Ramsey fringes. 
(b) Experimental measurements of $\delta f$, (c) experimental measurement of $\delta f_{cor}$, which is $\delta f$ corrected for the collisional frequency shift displayed by the solid red line of (a).}
\end{figure}

Following the previously described experimental protocol for atom detection, at each measurement cycle we detect both states $\left|b\right>$ and $\left|a\right>$, then for each image the atom number in each state -- $N_{\left|b\right>}$ and $N_{\left|a\right>}$ -- is computed by summing the pixels of the optical density over two regions of interest -- one for each state. 
Typically the region of interest for $\left|b\right>$ is 20 by 25 pixels and for $\left|a\right>$ is 40 by 35 pixels. 
The dimensions of these regions are optimised for maximizing the signal-to-noise ratio of the atom number. 
Then, we compute population in both states $P_{\left|b\right>}$ and $P_{\left|a\right>}$ by normalisation of the two previous atom numbers by the total atom number $N_{tot}=N_{\left|b\right>}+N_{\left|a\right>}$ \cite{Santarelli1999}.

For each image, we also measure the maximum of the optical density in state $\left|b\right>$ -- $\mathcal{O}^{max}_{\left|b\right>}$ -- by averaging the 40 maximum pixels of the optical density over the region of interest defined for the $N_{\left|b\right>}$ measurement (the same is done for state $\left|a\right>$). 
This maximum is inversely proportional to the atom cloud temperature $T$ \cite{Lye2002} - $\mathcal{O}^{max}_{\left|b\right>} \propto N_{\left|b\right>}/T$ and $\mathcal{O}^{max}_{\left|a\right>} \propto N_{\left|a\right>}/T$ and thus allows us to keep track of the atomic cloud temperature fluctuations during the Allan deviation measurement. 
For each image, from $P_{\left|b\right>}$ and $P_{\left|a\right>}$, we can compute $\delta f$ the frequency deviation using either of two following equations:
\begin{eqnarray}
\delta f & = & \frac{1}{2\pi T_R} \text{acos}\left( 1 - 2P_{\left|b\right>}  \right) \,, \nonumber \\
\delta f & = & \frac{1}{2\pi T_R} \text{acos}\left( 1 + 2P_{\left|a\right>}  \right)  \,.
\end{eqnarray}

These frequency deviations are displayed in figure \ref{fig_cor_density}.b. 
From $N_{tot}$, $\mathcal{O}^{max}_{\left|b\right>}$ and $\mathcal{O}^{max}_{\left|a\right>}$ (the latter having been corrected for the longer time of flight), we also compute the atomic cloud density $n$ up to a proportionality factor:
\begin{equation}
n \propto n^{est} = \frac{\left( \mathcal{O}^{max}_{\left|b\right>}+\mathcal{O}^{max}_{\left|a\right>} \right)^{3/2}}{\left( N_{tot} \right)^{1/2}}  \,.
\label{eq_DensityEstimate}
\end{equation}
The measured value of $\delta f$ as a function of $n^{est}$ (see figure \ref{fig_cor_density}.a) shows a strong linear correlation. 
As it will be explained with more details in section \ref{sec_ClockFreq}, this is the known collisional mean-field frequency shift $\delta f^{mean}_{col}$ \cite{Cutler2005,Laurent2020,Ovchinnikov2011,Wynands2005,Diddams2004,Abgrall2015,Bize2004,Guena2012}. 
One can show that \cite{Clairon1995,Walraven2010,Szmuk2015}:
\begin{eqnarray}
\delta f^{mean}_{col} & = & - \frac{2\hbar\left( a_{aa}-a_{bb} \right)}{m} \frac{1}{N} \int n^2(x,y,z) dx dy dz \label{eq_ColShift}\\ 
& = & - \frac{\hbar\left( a_{aa}-a_{bb} \right)}{\sqrt{2}m} n_0 \label{eq_ColShiftAverage}
\end{eqnarray}
where $m$ is the atomic mass, $a_{aa}$ and $a_{bb}$ are the $s$-wave scattering lengths between two atoms in state $\left|a\right>$ and $\left|b\right>$.
For rubidium 87 $a_{aa}=100.44a_0$, $a_{bb}=95.47a_0$, and $a_0 = 0.529\cdot10^{-10}$~m is the Bohr radius \cite{Egorov2013}. 
The density of a thermal cloud at equilibrium in a harmonic trap is given by \cite{Walraven2010}:
\begin{equation}
n(x,y,z) = n_0 \exp\left( -\frac{x^2m\omega_x^2}{2k_BT} -\frac{y^2m\omega_y^2}{2k_BT} -\frac{z^2m\omega_z^2}{2k_BT} \right)
\label{eq_CloudDensity}
\end{equation}
with the peak density:
\begin{equation}
n_0=N\omega_x\omega_y\omega_z\left( \frac{m}{2\pi k_B T} \right)^{3/2} \,.
\label{eq_PeakDensity}
\end{equation}
Thus we compute a new frequency deviation $\delta f_{cor}$ that is corrected for the collisional frequency shift:
\begin{equation}
\delta f_{cor} = \delta f - kn^{est}
\end{equation}
where $k$ is the fitted slope of the data in figure~\ref{fig_cor_density}.a (the red line). 
The corrected frequency deviations $\delta f_{cor}$ are displayed on figure \ref{fig_cor_density}.c and show a drastic reduction of their fluctuations from those in figure \ref{fig_cor_density}.b.

From this set of $\delta f_{cor}$ values, we compute the Allan deviation $\sigma(\tau)$ with overlapping samples \cite{Riley2008}:
\begin{equation}
\sigma^2(\tau = l\tau_0) = \frac{1}{2(N_s+1-2l)}\sum_{j=1}^{N_s-2l+1} \left[ \overline{y}_{j+l}^{l} - \overline{y}_j^{l} \right]^2
\end{equation}
where the $l$-th order mean of the sample $y_i$ is:
\begin{equation}
\overline{y}_{j+v}^{l} = \frac{1}{l} \sum_{u=j}^{l+j-1} y_{u+v}
\end{equation}
with $N_s$ the number of samples, $y_i=\delta f_{cor}^i/f^{mean}$, and $f^{mean}\approx6.834~678~116$~GHz the transition frequency. 
Two Allan deviations acquired on two different days \footnote{Although the data is a few years old at the time of this publication, it was chosen because it is the most comprehensive and complete data set we have for noise analysis, and other data collected is consistent with this noise analysis.} are displayed in figure \ref{fig_Allan_Result_1}.d. 
We obtain a stability at one shot (i.e., at 22~s) of $3.6\cdot 10^{-12}$ and $5.0\cdot 10^{-12}$ which integrates to $1\cdot 10^{-12}$ after 1000~s.

The values of $k$ found for the two previous datasets of figure \ref{fig_Allan_Result_1}.d differ by 20~$\%$, 
The variation of the value of $k$ means that there are other experimental correlations between the density $n$ and the frequency displacement $\delta f$ than the collisional frequency shift. 
This could lead to a shift on the correction of the frequency on the order of 20~$\%\times k \times <n^{est}>/f^{mean}\approx 5\cdot 10^{-12}$ which needs to be taken into account in the accuracy budget. 

\section{Choice of the magnetic field to operate the clock}
\label{sec_noise}

\begin{figure*}
\centering  \includegraphics[width=1.00\textwidth]{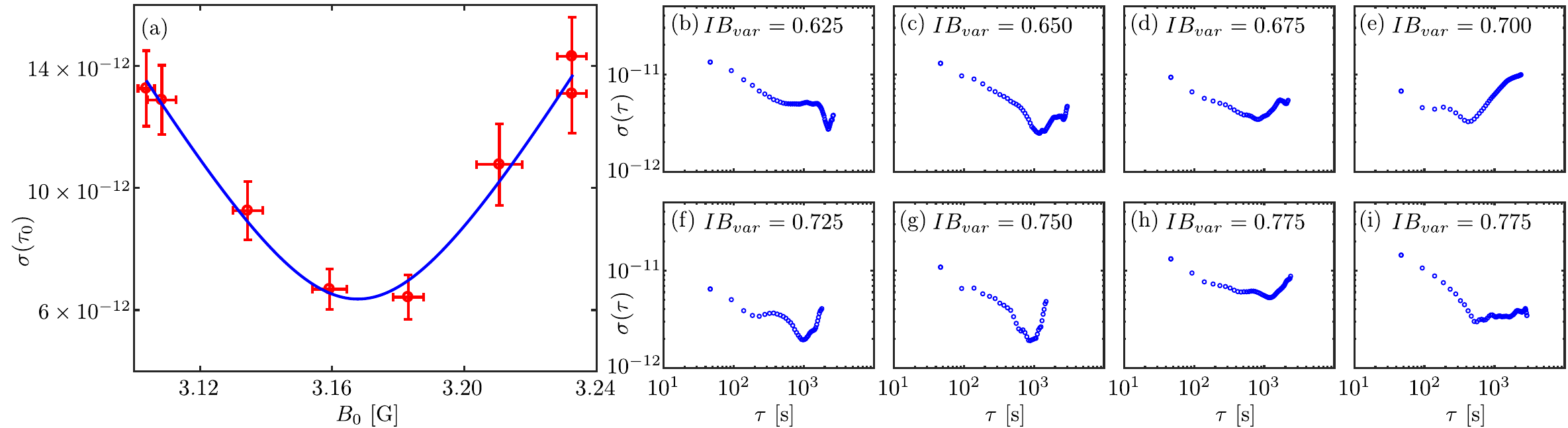}
\caption{\label{fig_AllanNoise} (Color online) Allan deviations recorded in the traps of figure \ref{fig_Trap_Bottom} using the Ramsey times computed from the data of figure \ref{fig_RamseyFringes}.
Those measurements lead to the computation of the impact of magnetic field and temperature noises on the stability of the atomic transition frequency (see section \ref{sec_ExpChooseMag}). 
(a) Allan deviation at $\tau_0$, $\sigma(\tau_0)$, of the relative transition frequency as a function of the magnetic field at the bottom of the trap $B_0$~[G]. 
The red open circles are experimental data and the solid blue line is a fit (see text). 
(b) to (i) Allan deviation, $\sigma(\tau)$, of the relative transition frequency as a function of the integration time $\tau$~[s] for several values of the $IB_{var}$ parameter (see text).}
\end{figure*}

In this section, we establish the dependence of the clock transition frequency with the magnetic field, the temperature and the atom number. 
Because the atomic cloud occupies a temperature dependent volume in the magnetic trap, the optimum magnetic field at the trap bottom does not correspond to the magic field and is dependent on both the magnetic field and the temperature fluctuations.
We will see below (section \ref{sec_RabiNoise}) that the contribution from the noise in the Rabi pulses also depends on the magnetic-field noise but that its contribution to the  noise budget of the clock is negligible. 
Thus, we do not consider it in the following optimization.

\subsection{Dependence of the transition frequency with magnetic field, atom number and temperature}
\label{sec_ClockFreq}

The transition frequency $f_t$ between $\left|a\right>$ and $\left|b\right>$ is given by:
\begin{eqnarray}
f_t & = & f_{\left|b\right>}^{BR} - f_{\left|a\right>}^{BR} + \frac{2\hbar (a_{bb}-a_{aa})}{m}n(x,y,z) \nonumber\\
& + & \frac{2\hbar}{m}(2a_{ab}-a_{aa}-a_{bb}) \nonumber\\
& & \times \left( n_{\left|a\right>}(x,y,z)-n_{\left|b\right>}(x,y,z) \right)
\end{eqnarray}
where $2\pi\hbar f_{\left|i\right>}^{BR}$ is the field dependent energy of state $\left|i\right>$ computed with the Breit-Rabi formula \cite{Steck2003} and $n_{\left|i\right>}(x,y,z)$ is the spatial density of atoms in state $\left|i\right>$. 
The last two terms are the mean-field shifts due to collisions and introduce a dependence of $f_t$ with the temperature $T$ and the atom number $N$. 
$a_{ab}=98.09a_0$ is the $s$-wave scattering length between an atom in state $\left|a\right>$ and one in $\left|b\right>$. 
We suppose that after the first $\pi/2$-pulse used for the Ramsey interferometer $n_{\left|a\right>}\approx n_{\left|b\right>}$ such that the last term of $f_t$ can be neglected.
We expand $f_{\left|b\right>}^{BR}$ and $f_{\left|a\right>}^{BR}$ up to the second order with $B$, and write the transition frequency as \cite{Rosenbusch2009,Szmuk2015}:
\begin{equation}
f= f_0+\delta f_B+\delta f_{col}
\end{equation}
with:
\begin{eqnarray}
f_0 & = & f_{hfs}+\frac{\mu_B g_I B^0_m}{2\pi\hbar} \\
\delta f_B & = & b \left(B(x,y,z)-B_m^0\right)^2 \\
\delta f_{col} & = & \frac{2\hbar (a_{bb}-a_{aa})}{m}n(x,y,z)
\end{eqnarray}
where $f_{hfs}$ is the frequency of the hyperfine splitting between state $F=1$ and $F=2$ of rubidium 87 in the absence of magnetic field, $\mu_B$ is Bohr magneton, $g_I=-9.95\cdot 10^{-4}$ is nuclear gyromagnetic factor, $B_m^0\approx3.229$~G is the magnetic magic field, $b\approx 431$~Hz/G$^2$, and $B(x,y,z)$ is an effective trapping magnetic field taking into account the whole potential seen by the atoms. 
$B(x,y,z)$ is equal to:
\begin{eqnarray}
B(x,y,z) & =&  B_0 + \frac{m\omega_x^2}{\mu_B}x^2 + \frac{m\omega_y^2}{\mu_B}y^2 + \frac{m\omega_z^2}{\mu_B}z^2 \nonumber\\
& & - \frac{2mg}{\mu_B}z
\end{eqnarray}
where $g\approx 9.81$~m$\cdot$s$^{-2}$ is the acceleration of gravity which shifts the minimum of the magnetic trap along the $z$ axis.

When measuring the fluctuation of $f_t$ with the Ramsey sequence, we measure the mean of the fluctuations of $f_t$ over space weighted by the atomic density. 
Computing this mean gives:
\begin{eqnarray}
f^{mean} & = & \frac{1}{N} \int f_t(x,y,z)n(x,y,z) dxdydy \\
& = & f_0 + \delta f_B^{mean} + \delta f_{col}^{mean}
\label{eq_FMean}
\end{eqnarray}
with $\delta f_{col}^{mean}$ given by the equation (\ref{eq_ColShiftAverage}) and $\delta f_B^{mean}$ given by:
\begin{eqnarray}
\delta f^{mean}_B & = & \frac{b}{\mu_B^2 } \left( \frac{4g^2 m k_B T}{\omega_z^2} +15k_B^2 T^2 \right. \nonumber\\
& & \left. +6\mu_B \delta B k_B T+\delta B^2 \mu_B^2 \right)
\end{eqnarray}
where $\delta B = B_0 - B^0_m$. 
As the Ramsey time $T_R=603.3$~ms is of the same order of magnitude as the trap lifetime $\tau_{trap}=1.2$~s, when evaluating $\delta f_{col}^{mean}$, we took for the number of atoms $N$ in equation (\ref{eq_ColShiftAverage}) the mean number of atoms in the trap during the Ramsey sequence $\overline{N}$.
This number is related to the number of atoms at the start of the Ramsey sequence $N_i$ by the formula:
\begin{equation}
\overline{N}=N_i \frac{1-e^{-T_R/\tau_{trap}}}{T_R / \tau_{trap}} \,.
\label{eq_MeanAtomNumber}
\end{equation}
Leading to $\overline{N}=(30.3 \pm 1.5)\cdot 10^{3}$ atoms.

The above equations show that $f^{mean}$ depends on the atom number, the temperature and the magnetic field. 
To lower the fluctuations of $f^{mean}$, we want to operate at a magnetic field minimizing the effect of the fluctuations of the magnetic field and of the temperature on the atomic transition frequency. 
The derivative of $f^{mean}$ with respect to $N$ is independent of the magnetic field so that the atom number fluctuations do not affect the choice of the magnetic field $B_0$. 

\subsection{Minimizing the effect of magnetic field noise}

Differentiation of the previous expression for $f^{mean}$, with respect to $\delta B$, leads to the following magnetic field which minimizes the fluctuations of $f^{mean}$ due to the magnetic-field fluctuations:
\begin{equation}
B_0^B=-\frac{3k_B T}{\mu_B}  + B_m^0 \,.
\end{equation}
If we chose a magnetic field $B_0$ not equal to $B_0^B$ but close to it, the relative stability of the frequency $f^{mean}$ is:
\begin{eqnarray}
\sigma_y^{Mag} & = & \frac{1}{f^{mean}} \left| \left. \frac{\partial f^{mean}}{\partial \delta B} \right|_{\delta B = B_0 - B_m^0} \right| \sigma_B \nonumber\\
& = & \frac{2b}{f^{mean}}\left| B_0-B_0^B \right| \sigma_B \,.
\label{eq_MagNoiseImpact}
\end{eqnarray}
Where $\sigma_B$ is the magnetic-field instability.

\subsection{Minimizing the effect of temperature noise}

Differentiation of $f^{mean}$ with respect to $T$, leads to the following magnetic field which minimizes the fluctuations of $f^{mean}$ with respect to the atomic cloud temperature:
\begin{eqnarray}
& B_0^T & = B_m^0 - \frac{5 k_B T}{\mu_B} - \frac{2g^2 m}{3\mu_B\omega_z^2} \nonumber\\ 
& - & \frac{\mu_B}{4b} \frac{\hbar\left( a_{aa}-a_{bb} \right)}{\sqrt{2}m} n_0(\overline{N})\frac{1}{k_B T}  
\end{eqnarray}
where $n_0(\overline{N})=2.9\cdot 10^{18}$ atoms/m$^3$ is the peak atom density (equation (\ref{eq_PeakDensity})) evaluated for the mean number of atoms in the trap during the Ramsey sequence (equation (\ref{eq_MeanAtomNumber})). 
If we choose a magnetic field $B_0$ not equal to $B_0^T$ but close to it, the relative stability of the frequency $f^{mean}$ is:
\begin{eqnarray}
\sigma_y^{Temp} & = & \frac{1}{f^{mean}} \left| \left. \frac{\partial f^{mean}}{\partial T} \right|_{\delta B = B_0 - B_m^0} \right| \sigma_T \nonumber\\
& = & \frac{6bk_B}{\mu_B f^{mean} } \left| B_0- B_0^T\right| \sigma_T  \,.
\label{eq_TempNoiseImpact}
\end{eqnarray}
Where $\sigma_T$ is the temperature instability.

\subsection{Experimental protocol to choose the magnetic field}
\label{sec_ExpChooseMag}

For our experimental parameters, we have $B_0^B=3.207$~G and $B_0^T=3.152$~G, thus we need to tune the magnetic field between these two values to minimize the fluctuation of $f^{mean}$. 
In view of the fact that the fluctuations of the clock frequency $f^{mean}$ are mainly given by the fluctuations of the temperature and the magnetic field (this will be confirmed by studying the noise budget in section \ref{sec_NoiseBudget}), we have: 
\begin{eqnarray}
\sigma_y & = & \left[ \left(\frac{2b}{f^{mean}}  \left|B_0-B_0^B\right| \sigma_B \right)^2 \right. \nonumber\\
& & \left. +\left(\frac{6bk_B}{\mu_B f^{mean} } \left|B_0-B_0^T\right| \sigma_T \right)^2 + \sigma_0^2 \right]^{1/2} 
\label{eq_MagAndTempNoise}
\end{eqnarray}
where $\sigma_0$ stands for the fluctuations which are independent of the magnetic field and the temperature. 
Thus, we measure this quantity as a function of the magnetic field. 

We now describe the experimental protocol to minimize equation (\ref{eq_MagAndTempNoise}) as a function of $B_0$: 

\paragraph{Magnetic field calibration: }We calibrate the magnetic field at the bottom of the trap as a function of $IB_{var}$ which is proportional to the current running through the coils generating the biais field $A_1 \propto IB_{var}$ (calibration are reported in appendix figure \ref{fig_Trap_Bottom}.a in appendix \ref{Annexe_A_DataSup}). 
In the range of $IB_{var}$ explored, we can change the magnetic field at the bottom of the trap by 0.15~G while the trap eigen frequencies and the trap distance to the atom chip change by less than 5\%. 
Thus, we suppose them to stay fixed. 
The magnetic field is measured by counting atoms in states $\left|2,2\right>$ and $\left|b\right>$ remaining in the trap as a function of the frequency of a radio-frequency field interacting with the atoms (see figures \ref{fig_Trap_Bottom}.b to \ref{fig_Trap_Bottom}.i). 
The resonances found are adjusted by Lorentzian curves to infer the resonance frequencies, after which the corresponding magnetic field is deduced by inverting the Breit-Rabi formula \cite{Steck2003} (see figure \ref{fig_Trap_Bottom}.a). 
Error bars on the magnetic field are inferred from statistical errors on the adjusted positions of the resonance curves.

\paragraph{Tuning the $\pi/2$ pulses: }For each value of the magnetic field, we look for the minimum of the two photon Rabi oscillation frequency as a function of the local oscillator frequency $f=f_{MW}+f_{RF}$. 
As in section \ref{sec_RamseyInter}, the frequency of the radio-frequency photon is fixed at $f_{RF}=1$~MHz. 
The frequencies of the Rabi oscillations are plotted in appendix figures \ref{fig_Rabi_Osc}.c to \ref{fig_Rabi_Osc}.i as a function of the frequency of the local oscillator for several magnetic fields and are adjusted to a parabola (see equation (\ref{eq_GeneRabiFreq})). 
The minima of the Rabi frequencies $\Omega_R$ are plotted in figure \ref{fig_Rabi_Osc}.b as a function of the magnetic field. 
The frequency of the local oscillator $f=f_{RF}+f_{MW}$ at the minimum of the Rabi oscillation frequencies is also plotted as a function of the magnetic field in appendix figure \ref{fig_Rabi_Osc}.a. The values of $f$ and $\Omega_R$ from appendix figures \ref{fig_Rabi_Osc}.a and \ref{fig_Rabi_Osc}.b give the parameters to drive the two photon $\pi/2$ pulses in the Ramsey interferometer. 

\paragraph{Determination of the Ramsey time: }With the parameters of the previous paragraph, we record Ramsey fringes between 0 and 600~ms for several values of the magnetic field (see figures \ref{fig_RamseyFringes}.b to \ref{fig_RamseyFringes}.h in appendix). 
The fringes are fitted with a sinusoid without damping, data are displayed in three windows (0 to 80~ms, 240 to 320~ms and 520 to 600~ms) but the data in these three windows are adjusted with the same sinusoid. 
The frequencies of the Ramsey fringes are displayed in figure \ref{fig_RamseyFringes}.a in appendix as a function of the magnetic field. 
This allows us, for several magnetic fields, to selected the closest Ramsey time to 600~ms to operate the interferometer at mid-fringe. 

\paragraph{Allan deviation: }With those selected Ramsey times, we measure the Allan deviation for several magnetic fields (see figures \ref{fig_AllanNoise}.b to \ref{fig_AllanNoise}.i). 
The first point of the Allan deviation as a function of the magnetic field is shown in figure \ref{fig_AllanNoise}.a. 
The data is adjusted with equation (\ref{eq_MagAndTempNoise}) to find the value of $\sigma_B=0.791\pm0.055$~mG and $\sigma_T=28.3\pm1.2$~nK, and $\sigma_y$ is minimum around 3.17~G slightly above the 3.160~G selected for studying the stability of the transition frequency in section \ref{sec_experiment}. 
Between 3.17~G and 3.16~G the stability varies by less than 5~\%, making the requirement on the exact value of the magnetic field not so tight.
In figure \ref{fig_AllanNoise}.a, $\sigma_y$ is higher than in figure \ref{fig_Allan_Result_1}.d, because here we did not apply the density frequency shift correction to the data.

We can make an alternative estimate of the noise on the magnetic field from the knowledge of the trap lifetime $\tau_{trap}=1.2$~s, the trap distance from chip $h\sim 100$~$\mu$m and the assumption that the trap lifetime is limited by heating from the noise on the currents generating the magnetic field. 
We start by deducing the spectral noise density $S_I$ (under the hypothesis of a white noise) on the power supplies used for the chip wires \cite{Henkel2003}:
\begin{equation}
S_I \sim \frac{1}{\tau_{trap}} \frac{h^2}{\mu^2} \frac{0.16\cdot 10^{-18}}{I}
\end{equation}
where $S_I$ is in A$^2$/Hz, $\tau_{trap}$ is in seconds, $h$ is in $\mu$m, the magnetic moment of the trapped atoms $\mu$ is in unit of Bohr magneton $\mu_B$ and $I$ the wire current is in A. 
We find $S_I \sim 1.9\cdot 10^{-14}$~A$^2$/Hz and $S_I \sim 5.2\cdot 10^{-14}$~A$^2$/Hz for both power supplies.
Given the 16~kHz bandwidth of the power supplies, we deduce a rms relative noise on the current of $6.3\cdot 10^{-5}$ and $2.8\cdot 10^{-4}$ respectively leading to a total noise on the magnetic field of 0.90~mG, in good agreement with the previous number. 

\section{Noise Budget}
\label{sec_NoiseBudget}

In this section, we estimate all the known sources of fluctuation of the clock frequency. 
They are of three kinds: i)~atomic transition fluctuations (see sections \ref{sec_MagTherNoise}, \ref{sec_QPNoise}, \ref{sec_RabiNoise}, \ref{sec_DensityCorrNoise} and \ref{sec_AtomLossNoise}), ii)~local oscillator noise (see section \ref{sec_DickNoise}) and iii)~technical detection noise (see section \ref{sec_DetNoise}).

\subsection{Magnetic and thermal noise}
\label{sec_MagTherNoise}

While looking for the magnetic field minimizing the clock frequency fluctuations, we computed the fluctuations of the magnetic field $\sigma_B$ and of the temperature $\sigma_T$. 
Using the value of $B_0^B$ and $B_0^T$ and equations (\ref{eq_MagNoiseImpact}) and (\ref{eq_TempNoiseImpact}), we compute their impact on the clock fluctuations. 
Thus for the magnetic field:
\begin{equation}
\sigma_y^{Mag} = 4.69\cdot 10^{-12}
\end{equation}
and for the temperature:
\begin{equation}
\sigma_y^{Temp} = 1.28\cdot 10^{-12} \,.
\end{equation}

\subsection{Quantum projection noise}
\label{sec_QPNoise}

We suppose the atoms are  uncorrelated, thus everything happens as if we are averaging $N$ uncorrelated measurements. 
The quantum projection noise gives the following noise on the population measurement \cite{Itano1993}:
\begin{equation}
\sigma_P=\frac{1}{2\sqrt{N}} \,.
\end{equation}
Using the population equations (for example in state $\left|b\right>$):
\begin{equation}
P_{\left|b\right>} = \frac{1}{2}\left[1-C\cos\left(\left(\omega-2\pi f^{mean}\right)T_R\right)\right]
\end{equation}
and the interrogation of the clock around mid-fringe (i.e., for a 50\% excitation probability), we can link the uncertainty on the population measurement to the uncertainty on the relative frequency:
\begin{equation}
\sigma_P = \left. \frac{\partial P}{\partial f^{mean}} \right|_{mid-fringe} \sigma_{f^{mean}}
\end{equation}
leading to a relative frequency uncertainty:
\begin{equation}
\sigma_y^{QP} = \frac{1}{f^{mean} \pi T_R C 2\sqrt{N}} \,,
\end{equation}
where $\sigma_y = \sigma_{f^{mean}}/f^{mean}$.
For the Ramsey time $T_R=603.3$~ms, the atom number at the end of the Ramsey $N_f=(23.4 \pm 1.2)\cdot 10^3$ and the contrast of $C=0.5$, we found:
\begin{equation}
\sigma_y^{QP} = 0.50\cdot 10^{-12} \,.
\end{equation}

\subsection{Technical detection noise}
\label{sec_DetNoise}

The population in state $\left|i\right>$ is $P_i = N_i/(N_a+N_b)$. 
The measurement of the atom number $N_i$ in state $\left|i\right>$
has a noise $\delta N_i$.
Thus, the noise on the population is:
\begin{equation}
\delta P_a = \frac{N-N_a}{N^2}\delta N_a - \frac{N_a}{N^2}\delta N_b = \frac{1}{2N}\left( \delta N_a - \delta N_b \right) \,.
\end{equation}
We assume the clock is interrogated at mid-fringe, thus $N_a/N^2=N_b/N^2 \approx 1/2N$. 
A similar equation holds for $\delta P_b$. 
In the experiment, we consider that technical detection noises $\delta N_a$ and $\delta N_b$ are uncorrelated thus:
\begin{equation}
\sigma_P=\sqrt{\left(\frac{\delta N_a}{2N}\right)^2 + \left(\frac{\delta N_b}{2N}\right)^2} \,.
\end{equation} 
And at mid-fringe, the noise on the relative frequency is:
\begin{equation}
\sigma_y^{Det} = \frac{1}{f^{mean} \pi T_R C}\sigma_P \,.
\end{equation}

To measure $\delta N_a$ and $\delta N_b$, we run around one hundred experimental cycles without loading atoms on the chip. 
On all the output optical density images, using the regions of interest for $\left|a\right>$ and $\left|b\right>$ used in the Allan deviation computation, we computed around one hundred \textquotedblleft atom number\textquotedblright \footnote{The measurement of $\delta N_a$ and $\delta N_b$ without loading atoms on the chip allows us to measure the technical detection noise without the contribution of the quantum projection noise from the atoms.}. 
Extracting the standard deviation of those data, we found $\delta N_a=136$~atoms and $\delta N_b=97$~atoms (ROIs are given in section \ref{sec_AllanVar}). 
$\delta N_a$ is greater than $\delta N_b$ because the time of flight for $\left|a\right>$ is greater than for $\left|b\right>$, leading to a lower signal to noise ratio on $N_a$ than on $N_b$. 
This measurement account for the noise of the camera and of the detection laser power and frequency noises. 
Taking into account the atom number at the end of the Ramsey sequence $N_f=(23.4\pm 1.2)\cdot 10^3$ and the numbers already used in section \ref{sec_QPNoise}, we find $\sigma_P=0.0036$ and:
\begin{equation}
\sigma_y^{Det} = 0.56 \cdot 10^{-12} \,.
\end{equation}
This value is slightly above the quantum projection noise, which prove the quality of the optical density imaging denoise procedure introduced in section \ref{sec_ODNoiseReduc}.

\subsection{Noise on the Rabi pulses}
\label{sec_RabiNoise}

The Rabi pulses used in the experiment are not exactly $\pi/2$ pulses, there are fluctuations $\delta\Omega$ and $\delta\tau_p$ of the generalized Rabi frequency $\Omega$ and Rabi time $\tau_p$. 
This lead to fluctuations of the clock populations:
\begin{equation}
\sigma_P = \frac{\pi}{4}\sqrt{ \left(\frac{\delta\Omega}{\Omega}\right)^2 + \left(\frac{\delta\tau_p}{\tau_p}\right)^2 } \,.
\end{equation} 
$\Omega$ depends on the RF and MW field amplitudes ($\Omega_{RF}$ and $\Omega_{MW}$), on the one and two photon detuning $\Delta$ and $\delta$ (see figure \ref{fig_RbLevel}), and on the light shift $\delta_{21}$. 
$\delta\Omega/\Omega$ can be written as a function of the noises on the experimental parameters:
\begin{eqnarray}
& & \left( \frac{\delta\Omega}{\Omega} \right)^2 = \left( \frac{\delta\Omega_{RF}}{\Omega_{RF}} \right)^2 \left( \frac{\Omega_R^2}{\Omega^2} \right)^2 \nonumber\\
& + & \left( \frac{\delta\Omega_{MW}}{\Omega_{MW}} \right)^2 \left( \frac{\Omega_R^2}{\Omega^2} - \frac{2\delta_{21}(\delta-\delta_{21})}{\Omega^2} \right)^2 \nonumber\\
& + & \left( \frac{\alpha \sigma_B}{\Delta} \right)^2 \left( -\frac{\Omega_R^2}{\Omega^2} + \frac{(\delta-\delta_{21})2b\delta B\Delta}{\Omega^2\alpha} \right. \nonumber\\
& & \quad \quad \quad \quad \left. + \frac{\delta_{21}(\delta-\delta_{21})}{\Omega^2} \right)^2 \nonumber\\
& + & \left( \frac{\delta\omega_{MW}}{\Delta} \right)^2 \left( -\frac{\Omega_R^2}{\Omega^2} - \frac{\Delta(\delta-\delta_{21})}{\Omega^2} + \frac{\delta_{21}(\delta-\delta_{21})}{\Omega^2} \right)^2 \nonumber\\
& + & \left( \frac{\delta\omega_{RF}}{\Delta} \right)^2 \left( \frac{\Delta(\delta-\delta_{21})}{\Omega^2} \right)^2
\end{eqnarray}
where $\delta\Omega_{RF}$, $\delta\Omega_{MW}$, $\delta\omega_{MW}$ and $\delta\omega_{RF}$ are respectively the noise on the radio and microwave amplitudes, and the noise on the microwave and radio frequencies. 
Taking into account our experimental parameters: $\Omega\approx\Omega_R\approx 250$~Hz, $\delta_{21}\approx92$~Hz, $\delta-\delta_{21}\approx3$~Hz, $\Delta\approx1.209$~MHz and $\alpha\approx0.7$~MHz/G we found that the main contribution are given by:
\begin{eqnarray}
\left( \frac{\delta\Omega}{\Omega} \right)^2 & = &
\left( \frac{\delta\Omega_{RF}}{\Omega_{RF}} \right)^2 + \left( \frac{\delta\Omega_{MW}}{\Omega_{MW}} \right)^2 + \left( \frac{\alpha \sigma_B}{\Delta} \right)^2 \nonumber\\
& + & 60\times  \left( \frac{\delta\omega_{MW}}{\Delta} \right)^2 + 60\times  \left( \frac{\delta\omega_{RF}}{\Delta} \right)^2 \,.
\end{eqnarray}
In our experimental setup we estimate $\delta\Omega_{RF}/\Omega_{RF}$ and $\delta\Omega_{MW}/\Omega_{MW}$ to be equal to $10^{-4}$, $\alpha\sigma_B/\Delta=4.58\cdot 10^{-4}$, and $\delta\omega_{RF}/\Delta$ and $\delta\omega_{MW}/\Delta$ are smaller than $10^{-7}$ thus these last two terms do not contribute to the noise on the Rabi pulses.
This leads to\footnote{We considered that $\delta\tau_p/\tau_p$ is negligeable in our experiment.}:
\begin{equation}
\sigma_y^{Rabi} = 0.06 \cdot 10^{-12} \,.
\end{equation}
As stated above, the fluctuations of the Ramsey pulses make negligible contributions to our noise budget.

\subsection{Noise on the density correction}
\label{sec_DensityCorrNoise}

As explained in section \ref{sec_AllanVar}, the atomic density fluctuates from shot to shot, thus we corrected the measured frequency by $k\times n^{est}$, where $k\approx - 17$~Hz/a.u. is the slope of the linear fit of the figure \ref{fig_cor_density}.a (the a.u. arbitrary units are the same as in the figure \ref{fig_cor_density}) and $n^{est}$ is given by equation (\ref{eq_DensityEstimate}). 
The noise on this correction $\sigma_y^{DensCorr}$ depends on the noise $\delta k$ on the slope of the density correction $k$, on the noise in the total dected atom number $\delta N_{tot}$ and on the measurement of the maximum of the optical density noise $\delta \mathcal{O}^{max}$. 
This leads to:
\begin{eqnarray}
\sigma_y^{DensCorr} & = & \frac{1}{f^{mean}} \left[ \left( \delta k\frac{ \left(\mathcal{O}^{max}\right)^{3/2}}{\left(N_{tot}\right)^{1/2}} \right)^2 \right. \nonumber\\
& + & \left( \frac{3}{2} k \frac{ \left(\mathcal{O}^{max}\right)^{1/2}}{\left(N_{tot}\right)^{1/2}} \delta \mathcal{O}^{max} \right)^2 \nonumber\\
& + & \left. \left( \frac{1}{2} k \frac{ \left(\mathcal{O}^{max}\right)^{3/2}}{\left(N_{tot}\right)^{3/2}} \delta N_{tot} \right)^2 \right]^{1/2}
\label{eq_ErrorDensCor}
\end{eqnarray}
$\delta k$ is estimated to 1.8\% of $k$, the noise on the total detected atom number is given in section \ref{sec_DetNoise}. 
In the (arbitrary) units used here we have $N_{tot}=210$ and $\delta N_{tot}=1.7$. 
The noise on the measurement of the optical density is estimated using the optical density images without atoms described in the previous section. 
Using the same a.u. unit we have $\mathcal{O}^{max}=0.35$ and $\delta \mathcal{O}^{max}=0.006$. 
This leads to:
\begin{equation}
\sigma_y^{DensCorr} = 1.13 \cdot 10^{-12}
\end{equation}
where the main contribution comes from the measurement of the maximum of the optical density.

\subsection{Noise on the atomic losses}
\label{sec_AtomLossNoise}

The trap lifetime is not long compared with the interrogation time. 
Because of this the atom density and therefore the density correction decreases with time. 
Since the atom loss is stochastic, the average density correction is subject to noise. 
Knowing the trap lifetime $\tau_{trap}$ and the number of atoms at the end of the Ramsey sequence $N_{\mathrm{f}}$, it can be shown that the uncertainty in the atom number $N_t$ at a time $t$ during the Ramsey sequence is \cite{Szmuk2015}:
\begin{equation}
\sigma_{N_t}(t) = \sqrt{\left(1-e^{\frac{T_R-t}{\tau_{trap}}}\right)\left(2-\left(N_{\mathrm{f}}+2\right)e^{\frac{T_R-t}{\tau_{trap}}}\right)} \,.
\end{equation}
Integrating this expression over $t$ between $0$ and $T_R$, gives the mean error on the atom number $\sigma_{\overline{N}}=(1/T_R)\int^{T_R}_0\sigma_{N_t}(t)dt $. 
Last term of equation (\ref{eq_ErrorDensCor}) gives the impact of $\sigma_{\overline{N}}$ on the relative frequency stability of the clock:
\begin{equation}
\sigma_y^{Losses} = \frac{k}{2f^{mean}}\frac{\left(\mathcal{O}^{max}\right)^{3/2}}{\left(N_{tot}\right)^{1/2}} \frac{\sigma_{\overline{N}}}{\overline{N}} \,.
\end{equation}
$\sigma_{\overline{N}}=91$~atoms and the numbers of the previous sections give:
\begin{equation}
\sigma_y^{Losses} = 0.05\cdot 10^{-12} \,.
\end{equation}

\subsection{Dick effect}
\label{sec_DickNoise}

The Ramsey measurement amounts to periodically sampling the local oscillator frequency and its fluctuations: this leads to aliasing. 
Thus, high-frequency noise of the local oscillator close to multiples of the sample frequency degrades the long term stability of the clock. 
This is known as the Dick effect, which gives the following relative frequency instability of the clock \cite{Szmuk2015,Szmuk2015b,Dick1989,Santarelli1998}:
\begin{equation}
\sigma_y^{Dick} = \left[ \frac{1}{\tau_0} \sum_{l=1}^\infty \left(\frac{g_l}{g_0}\right)^2 S_y^f\left(\frac{l}{\tau_0}\right) \right]^{1/2}
\end{equation}
where $\tau_0=22$~s is the duty cycle of the clock, and $S_y^f (f)$ is the power spectral density of the local oscillator frequency noise, and:
\begin{equation}
g_l=\frac{1}{\tau_0} \int_{-\tau_0/2}^{+\tau_0/2} g(t) \cos\left(\frac{2\pi l t}{\tau_0}\right)dt
\end{equation}
with $g(t)$ the sensitivity function of the interferometer. If we operate at mid-fringe with $\left|\delta\right| \ll \Omega_R$, $g(t)$ is given by \cite{Santarelli1998}:
\begin{eqnarray}
g(t) & = & \sin\left[\Omega_R\left(T_R/2+\tau_p+t\right)\right] \nonumber\\ 
& \text{if} & \qquad -\tau_p-T_R/2\leq t \leq -T_R/2 \nonumber\\
g(t) & = & \sin\left[\Omega_R \tau_p \right] \nonumber\\ 
& \text{if} & \qquad -T_R/2 \leq t \leq T_R/2 \nonumber\\
g(t) & = & \sin\left[\Omega_R\left(T_R/2+\tau_p-t\right)\right] \nonumber\\
& \text{if} & \qquad T_R/2 \leq t \leq T_R/2+\tau_p \nonumber\\
g(t) & = & 0 \qquad \text{otherwise.}
\end{eqnarray}
Where $\tau_p$ is the $\pi/2$ pulse duration and $\Omega_R$ the Rabi frequency.

The Allan deviation of our local oscillator has been measured to be
\footnote{Our local oscillator is a GNSS disciplined quartz. We know that around 1~s the relative frequency stability of the quartz is about $2.5\cdot10^{-13}$, which is below the value given in equation (\ref{eq_LocOscAllan}), and that the time constant of the lock on GNSS is around $10^4$~s. }:
\begin{equation}
\sigma_y^{osc} = \sqrt{ \left( 5\cdot 10^{-12} / \tau \right)^2 + \left( 4\cdot 10^{-13} \right)^2 }
\label{eq_LocOscAllan}
\end{equation}
translating this into $S_y^f$ allows us to compute the contribution to the relative frequency instability of the clock:
\begin{equation}
\sigma_y^{Dick} = 0.72\cdot 10^{-12} \,.
\end{equation}

\subsection{Sumary of the noise budget}

We summarize all the noise sources in the second column of the table \ref{tab_NoiseSum}. 
The total estimated noise is $5.1\cdot 10^{-12}$ at one shot, slightly higher than the measured values, which are between $3.6\cdot 10^{-12}$ and $5\cdot 10^{-12}$. 
We can account for the discrepancy by examining correlations between the magnetic and temperature fluctuations. 
When we measured $\sigma_B$ and $\sigma_T$  using the Allan deviations of figure \ref{fig_AllanNoise}, the density correction of section \ref{sec_AllanVar} was not applied in the computation of the Allan deviations of figures \ref{fig_AllanNoise}.b to \ref{fig_AllanNoise}.i. 
But the density-fluctuation corrections to the raw data in section \ref{sec_AllanVar} also partially correct for the temperature and magnetic fluctuations. 
The temperature fluctuations are partially corrected because the density depends on the temperature. 
The magnetic-field fluctuations are also partially corrected because the cooling process (a RF knife in a magnetic trap) induces a strong correlation between the temperature and magnetic-field fluctuations. 
In addition, since the density correction corrects some part of the magnetic fluctuations, the Rabi fluctuations are also reduced.

\begin{table}
\caption{\label{tab_NoiseSum} Summary of the different contributions to the clock frequency fluctuations. Type (i) is atomic frequency noise, (ii) is local oscillator noise and (iii) is technical detection noise. The second column uses an estimation of the magnetic and thermal noises without the density correction, the third one gives it with the density correction. For the total, we considered that the noises are not correlated.}
\begin{ruledtabular}
\begin{tabular}{rrrl}
 Noise & $\times 10^{-13}$ & $\times 10^{-13}$ & type $\qquad$ \\
\hline
Magnetic & 46.9 & 34.3 & i \\
Thermal & 12.8 & 4.5 & i \\
Density correction & 11.3 & 11.3 & i \\
Dick effect & 7.2 & 7.2 & ii \\
Technical detection & 5.6 & 5.6 & iii \\
Quantum projection noise & 5.0 & 5.0 & i \\
Rabi fluctuations & 0.6 & 0.5 & i \\
Atomic losses & 0.5 & 0.5 & i \\
\hline
Total & 51.0 & 37.9 &  \\
\end{tabular}
\end{ruledtabular}
\end{table}

If we fit the value of $\sigma_T$ and $\sigma_B$ to the same data as in figure \ref{fig_AllanNoise}.a while using the density correction while computing the Allan deviations, we find $\sigma_T=10.0\pm 1.8$~nK and $\sigma_B=0.579\pm 0.081$~mG. 
These values lead to the noise levels given in the third column of the table~\ref{tab_NoiseSum}: $3.8\cdot 10^{-12}$ at one shot in better agreement with the measured values.

\section{Reduction of the clock frequency fluctuations}
\label{sec_discussion}

In this section, we first present a procedure to enhance the correction of the clock frequency displacement $\delta f$ with the total number of atoms $N_{tot}$ and the cloud temperature $T$. 
Then, we explore perspectives for further improvement of the Allan deviation.  

\begin{figure*}
\centering  \includegraphics[width=1.00\textwidth]{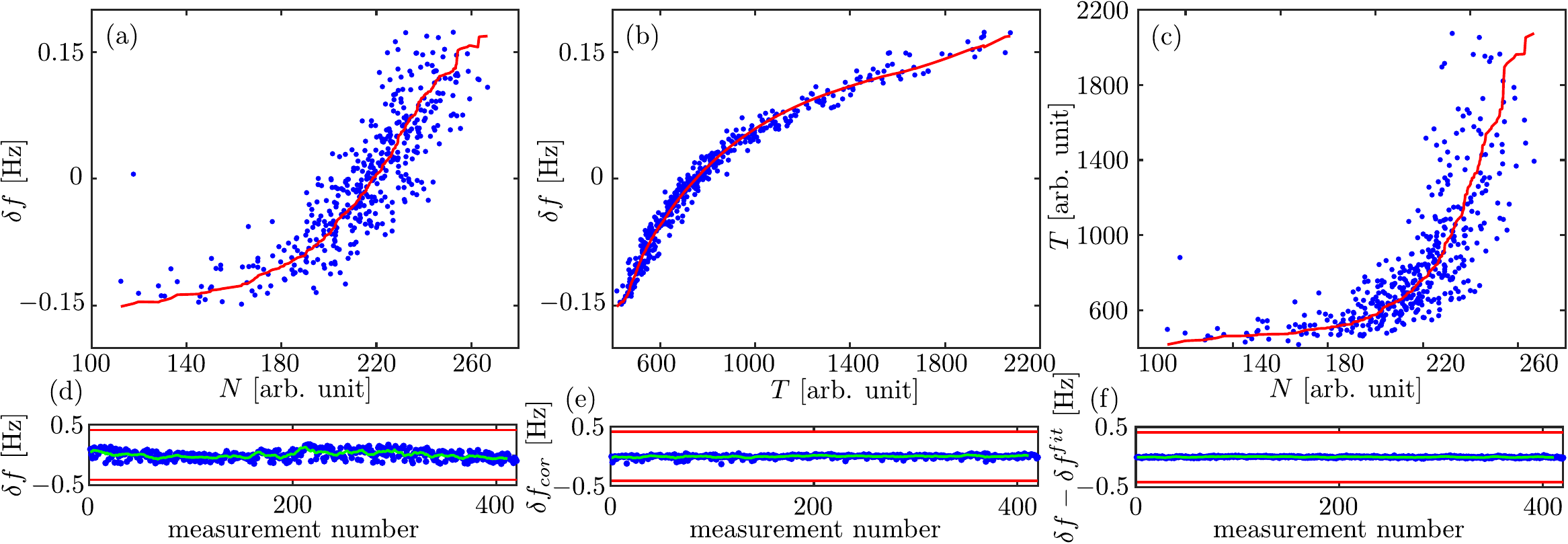}
\caption{\label{fig_cor_N_T} (Color online) Correlations between the frequency displacement, the atom number and the temperature used to improve (compare with figure \ref{fig_cor_density}) the correction of the frequency displacement with experimental parameter fluctuations. 
(a) Frequency displacement $\delta f$~[Hz] as a function of the atom number $N$~[arb. unit]. 
(b) Frequency displacement $\delta f$~[Hz] as a function of the temperature $T$~[arb. unit]. 
(c) Temperature $T$~[arb. unit] as a function of the atom number $N$ [arb. unit]. 
The blue dots are the data and the red line is a fit (see text). 
(d), (e) and (f), frequency displacement $\delta f$~[Hz] as a function of the measurement number. 
(d) Raw data. 
(e) Frequency displacement corrected with the process described in section \ref{sec_AllanVar}. 
(f) Frequency displacement corrected using the correlations displayed in (a), (b) and (c), (see text).}
\end{figure*}

\begin{figure*}
\centering  \includegraphics[width=0.85\textwidth]{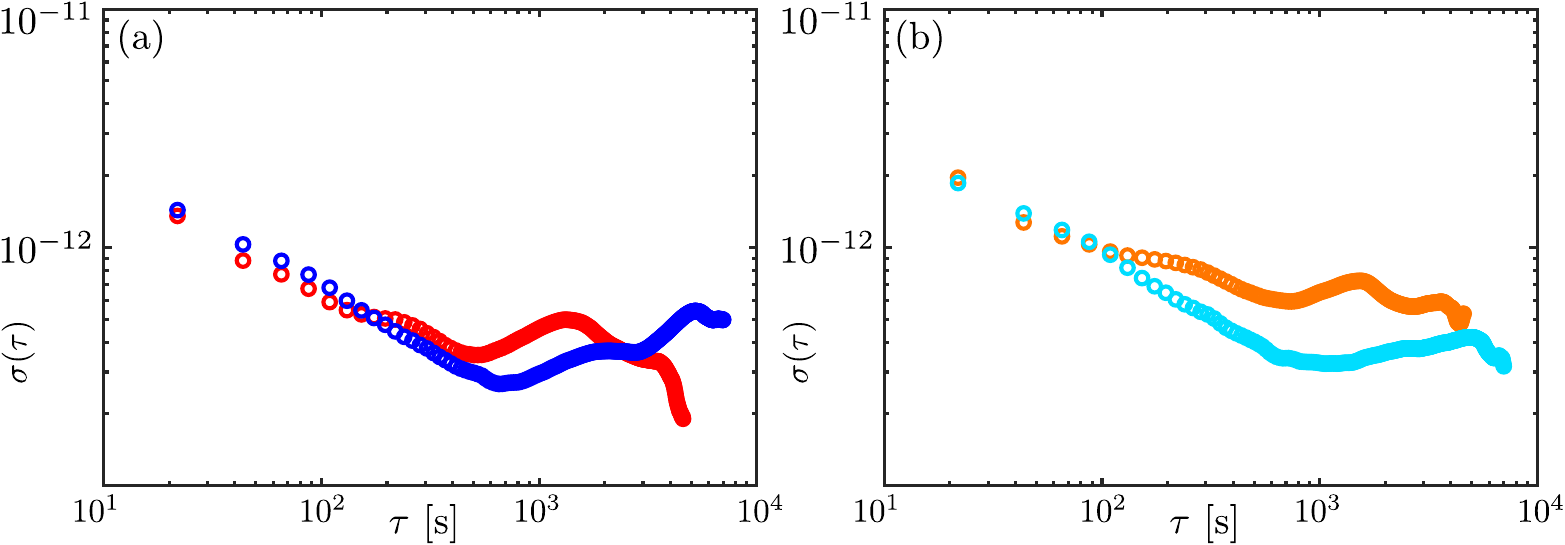}
\caption{\label{fig_Allan_Result_2} (Color online) (a) Allan deviation $\sigma(\tau)$ as a function of the integration time $\tau$~[s] after correction for the atom number and temperature fluctuations given by equation (\ref{eq_SecondCorr}), same raw data than in figure \ref{fig_Allan_Result_1}.d, in blue $1.4\cdot 10^{-12}$ (22$^{nd}$ November 2016), in red $1.3\cdot 10^{-12}$ (15$^{th}$ November 2016). 
(b) Allan deviation $\sigma(\tau)$ as a function of the integration time $\tau$~[s] after correction for the atom number and temperature fluctuations given by equation (\ref{eq_SecondCorr}). 
Light blue (respectively orange) points are the blue (respectively red) data of panel (a) but corrected with the coefficient of equation (\ref{eq_SecondCorr}) fitted on the red (respectively blue) data of panel (a). 
In light blue first point is $1.8\cdot 10^{-12}$, in orange $1.9\cdot 10^{-12}$. }
\end{figure*}

\subsection{A better correction of the density shift}

Figures \ref{fig_cor_N_T}.a to \ref{fig_cor_N_T}.c show the correlations between the frequency displacement $\delta f$, and the atom number $N_{tot}$ and the temperature $T$, as well as the correlation, which we have already mentioned, between the atom number and the temperature.
The raw data $\delta f$ can be fitted with a third-order polynomial in $N_{tot}$ and $T$:
\begin{equation}
\delta f^{fit} = \sum_{k=0}^3 \sum_{j=0}^k a_{j,k}N_{tot}^{k-j}T^j \,.
\label{eq_SecondCorr}
\end{equation}
This allows us to go beyond the correlation between $\delta f$ and the atom density and to take into account correlations that are present in the experiment like the one between the temperature and the magnetic field. 
This function is chosen empirically to fit the data of figure \ref{fig_cor_N_T}. 
Using this $\delta f^{fit}$ as a correction on the raw data $\delta f$ reduces the shot to shot frequency displacement as shown in figures \ref{fig_cor_N_T}.d to \ref{fig_cor_N_T}.f. 
We can use this correction to compute the Allan deviation displayed in figure \ref{fig_Allan_Result_2}.a with a stability between $1.3\cdot 10^{-12}$ and $1.4\cdot 10^{-12}$ at one shot which integrated to a few $10^{-13}$ in about 600~s where we reach the noise floor of our local oscillator (\ref{eq_LocOscAllan}) used for the characterization of the clock fluctuations.
With this correction, the difference between the first point of the Allan deviation of the two datasets is reduced from 40\% (figure \ref{fig_Allan_Result_1}.d) to 8\% (figure \ref{fig_Allan_Result_2}.a). 
To access this correction, we also corrected one dataset with the $a_{j,k}$ coefficients of equation (\ref{eq_SecondCorr}) ajusted on the other dataset. 
This result in the Allan deviation shown in figure \ref{fig_Allan_Result_2}.b reaching stabilities between $1.8\cdot 10^{-12}$ and $1.9\cdot 10^{-12}$. 

\subsection{Perspective for further frequency fluctuations reduction}

Let us suppose that we want to reach a total noise budget for the clock of few $10^{-13}$ at one shot and explore which improvement of the experimental parameters this implies. 

\paragraph{Quantum projection noise: }First of all, this requires increasing the Ramsey interrogation time by a factor of ten (from 0.6~s to 6~s). 
This can be achieved thanks to identical spin rephasing (while keeping a contrast above 0.5) \cite{Deutsch2012} as has already been used in the clock of reference \cite{Szmuk2015}. 
Such a Ramsey time also requires increasing the trap lifetime by a factor of ten.
Keeping the same trap parameters and using the atom heating model presented in the last paragraph of section \ref{sec_ExpChooseMag}, this also requires power supplies with a relative current stability of at least $2.8\cdot 10^{-4}\times \sqrt{1.2/12}\approx 9\cdot 10^{-5}$ (assuming the same bandwidth). 
These improvements lead to
\begin{equation}
\sigma_y^{QP} = 0.5\cdot 10^{-13} \,.
\end{equation} 

\paragraph{Detection noise: }The improvement of the Ramsey time also leads to a decrease of the technical detection noise to:
\begin{equation}
\sigma_y^{Det} = 0.6\cdot 10^{-13} \,.
\end{equation} 

\paragraph{Magnetic and thermal noises: }Improving the stability of the power supplies to the previously given number will not be enough to reduce the magnetic-field noise below $10^{-13}$.  
For achieveing this number, a magnetic-field noise below 15~$\mu$G must be reached. 
For 3.160~G, this can be done using power supplies for the two dimple wires with a relative current noise of $4.7\cdot 10^{-6}$ (the considered bandwidth is 16~kHz). 
Since the magnetic-field noise and the temperature noise are strongly correlated and proportional to each other, we also find a reduced temperature noise (10~nK $\times$ 15~$\mu$G/579~$\mu$G). 
This leads to:
\begin{equation}
\sigma_y^{Mag} = 0.9\cdot 10^{-13}
\end{equation} 
and:
\begin{equation}
\sigma_y^{Temp} = 0.1\cdot 10^{-13} \,.
\end{equation}

\paragraph{Rabi fluctuation noise: }Improving the radio frequency and microwave supplies to reach an amplitude stability of $\delta\Omega_{RF}/\Omega_{RF}=10^{-5}$ and $\delta\Omega_{MW}/\Omega_{MW}=10^{-5}$ would reduce the contribution of the noise on the Rabi pulses to:
\begin{equation}
\sigma_y^{Rabi} = 0.1\cdot 10^{-13} \,.
\end{equation}

\paragraph{Dick effect: }If the local oscillator and the total cycle time are kept unchanged, the Dick effect is reduced to $0.34\cdot 10^{-12}$, to keep it below $10^{-13}$ the total cycle time must be lowered to 7.3~s, leading to:
\begin{equation}
\sigma_y^{Dick} = 1.0\cdot 10^{-13} \,.
\end{equation}
Lowering the total cycle time to 7.3~s with a 6~s Ramsey time, leaves less than 1.3~s to cool atoms, compare with our actual 2.9~s cooling time. This can be implemented on an atom chip as described in reference \cite{Rudolph2015}. 

\paragraph{Atomic losses noise: }For a trap lifetime of $\tau_{trap}=12$~s and a Ramsey Time $T_R=6$~s the noise on the atomic losses is same as previously:
\begin{equation}
\sigma_y^{Losses} = 0.5 \cdot 10^{-13} \,.
\end{equation}

\paragraph{Density correction noise: }It remains to improve $\sigma_y^{DensCorr}$ which is link to the improvement of $\delta N_{tot}$, $\delta \mathcal{O}^{max}$ and $\delta k$ (see equation (\ref{eq_ErrorDensCor})).

\paragraph*{Description of the technical detection noise: }If the detection is limited by the poisionian distribution of the noise on the photoelectrons generated by the camera, it can be shown that the noise on the measured atom number per pixel $\delta N(i,j)$ is given by \cite{Maussang2010}:
\begin{eqnarray}
\delta N(i,j) & = & gL^2\left(\frac{1}{\left<I_A(i,j)\right>}+\frac{1}{\left<I_W^{cor}(i,j)\right>}\right) \nonumber\\
& + & gH^2\left( \left<I_A(i,j)\right> + \left<I_W^{cor}(i,j)\right>\right)
\end{eqnarray}
with:
\begin{eqnarray}
L=\frac{A}{\sigma_0} \qquad H = \frac{2}{\Gamma\tau_{img} g \eta T_{opt}}
\end{eqnarray}
where $g=2.6$ is the camera gain in term of count number per photon, $\left<I_A(i,j)\right>$ (respectively $\left<I_W^{cor}(i,j)\right>$) is the mean number of counts per pixels on a set of images with atoms $I_A$ (respectively without atoms $I_W^{cor}$). 
For our experimental parameters both are on the order of $2\cdot 10^4$ counts per pixel. 
$A=~3.76\times 3.76$~$\mu$m$^2$ is the pixel area in the object space, $\sigma_0=2.9\cdot 10^{-13}$~m$^2$ (respectively $\Gamma = 2\pi\times 6$~MHz) is the effective absorption area (respectively the natural linewidth) of the used detection transition, $\tau_{img}=35$~$\mu$s is the detection pulse duration, $\eta= 0.91$ is the quantum efficiency at 780~nm of the camera, and $T_{opt}=0.88$ is the transmission of the detection optic. 

\paragraph*{$\delta N_{tot}$: }All previous numbers lead to $\delta N(i,j) = 0.82 $~atom/pix, thus considering that the noises on different pixels are not correlated, this leads to 30.7 atoms for the ROI used to detect state $\left|a\right>$ and 18.3 atoms for the ROI used for $\left|b\right>$. 
This will lead to a reduction of $\delta N_{tot}/N_{tot}$ from 0.008 to 0.0015 and a reduction of the third term of equation (\ref{eq_ErrorDensCor}) from $1.44\cdot 10^{-13}$ to $0.27\cdot 10^{-13}$. This improvement will also lead to a further decrease of the technical detection noise to $\sigma_y^{Det} = 0.1 \cdot 10^{-13}$.

\paragraph*{$\delta \mathcal{O}^{max}$: }The noise per pixel on the optical density $\delta \mathcal{O}(i,j)$ can be deduced from the noise of the atom number per pixel $\delta N(i,j)$: $\delta \mathcal{O}(i,j)= \sigma_0 \delta N(i,j) /A$ leading to the following noise level on the maximum of the optical density $\mathcal{O}^{max}$: $\delta \mathcal{O}^{max}=\delta \mathcal{O}(i,j)/\sqrt{N_{pix}}=0.0168/\sqrt{N_{pix}}$, where $N_{pix}$ is number of pixel used in the average. 
Increasing this number of pixels from 40 (used to compute $\mathcal{O}^{max}$ in the density correction of figure \ref{fig_cor_density}) to 500 will lead to a decrease of the second term of equation (\ref{eq_ErrorDensCor}) from $9.14\cdot 10^{-13}$ to $2.58\cdot 10^{-13}$. 

\paragraph*{$\delta k$:} The noise on the slope of the density correction $\delta k$ is given by \cite{Johnston1997} $\delta k =\sigma_{R(\delta f)} / (\sqrt{N_{pt}} \sigma_{n})$, where $\sigma_{R(\delta f)}$ (respectively $\sigma_{n}$) is the standard deviation of the residual of the fit (along the horizontal axis of the data points in figure \ref{fig_cor_density}.a) and $N_{pt}$ is number of points. 
To reduce the amplitude of the first term of equation (\ref{eq_ErrorDensCor}) from $6.40\cdot 10^{-13}$ to $0.92\cdot 10^{-13}$, one needs to decrease $\delta k$ by a factor of 7. 
This can be done by increasing $N_{pt}$ from 420 to 2~940, with a clock duty cycle of 7.3~s this lead to 6 hours of data. 

All lead to the following noise on the density correction:
\begin{equation}
\sigma_y^{DensCorr} = 2.8\cdot 10^{-13} \,.
\end{equation}

The technical detection and the density correction noises could be further reduced by optimizing the time of flight. 
It will result in an increase of the ROIs for computing the atom number thus $\delta N_{tot}/N_{tot}$ will increase but this will also allow us to compute the parameter $\mathcal{O}^{max}$ over a higher number of pixel leading to reduction of $\delta \mathcal{O}^{max}$.

For all those improvements we used numbers that are reachable with a proper experimental design. This would lead to a total noise level of: $3.2\cdot 10^{-13}$ largely dominated by the density correction noise. 

\section{Conclusion}

We have demonstrated a cold atom clock on a chip with a noise (single shot stability) level between $3.6\cdot 10^{-12}$ and $5\cdot 10^{-12}$, we gave analytical model and experimental measurements of all the known noises of this clock that are in agreement with the measured Allan deviation. 
We also demonstrated that taking into account experimental correlation between the clock frequency displacement and the atom number and temperature, we can further reduce the clock noise to $1.4\cdot 10^{-12}$. 
We argue that there is substantial room for improvement in the magnetic field and detection noise which may allow us to reach a stability close to a few times $10^{-13}$.

\vspace*{0.3cm}
\begin{acknowledgments}
This work has been carried out within the ONACIS project ANR-13-ASTR-0031 and NIARCOS project ANR-18-ASMA-0007-02 funded by the French National Research Agency (ANR) in the frame of its 2013 Astrid and 2018 Astrid Maturation programs. 
This work also received funding from the European Defence Fund (EDF) under grant agreeement 101103417 - project ADEQUADE.
\end{acknowledgments}

%%%%%%%%%%%%%%%%%%%%%%%%%%%%%%%%%%%%%%%%%%%%%%%%%%%%%%%%%%
%%%%%%%%%%%%%%%%%%%%%%%%%%%%%%%%%%%%%%%%%%%%%%%%%%%%%%%%%%
%%%%%%%%%%%%%%%%%%%%%%%%%%%%%%%%%%%%%%%%%%%%%%%%%%%%%%%%%%
%%%%%%%%%%%%%%%%%%%%%%%%%%%%%%%%%%%%%%%%%%%%%%%%%%%%%%%%%%
%%%%%%%%%%%%%%%%%%%%%%%%%%%%%%%%%%%%%%%%%%%%%%%%%%%%%%%%%%
%%%%%%%%%%%%%%%%%%%%%%%%%%%%%%%%%%%%%%%%%%%%%%%%%%%%%%%%%%
%%%%%%%%%%%%%%%%%%%%%%%%%%%%%%%%%%%%%%%%%%%%%%%%%%%%%%%%%%
%%%%%%%%%%%%%%%%%%%%%%%%%%%%%%%%%%%%%%%%%%%%%%%%%%%%%%%%%%
%%%%%%%%%%%%%%%%%%%%%%%%%%%%%%%%%%%%%%%%%%%%%%%%%%%%%%%%%%
%%%%%%%%%%%%%%%%%%%%%%%%%%%%%%%%%%%%%%%%%%%%%%%%%%%%%%%%%%

\appendix

\section{Supplementary experimental data}
\label{Annexe_A_DataSup}

For sake of exhaustiveness, in this appendix, we gathered the data that are not necessary for a first reading of this paper but needed for a full analysis of our clock.

\begin{figure*}
\centering  \includegraphics[width=1.00\textwidth]{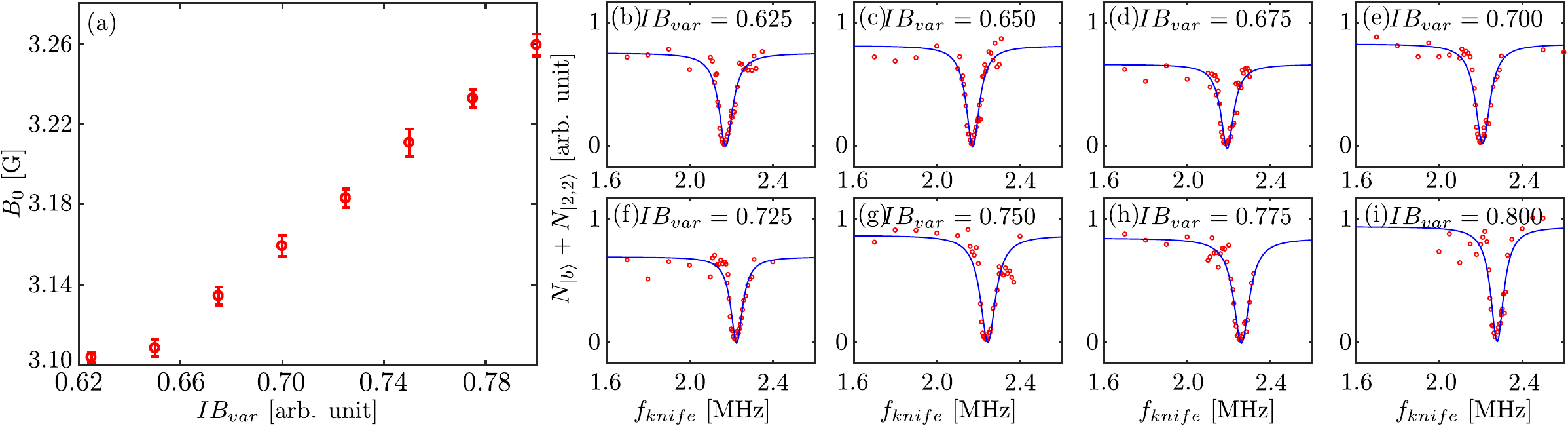}
\caption{\label{fig_Trap_Bottom} (Color online) Calibration of the magnetic field at the bottom of the trap for several traps. 
(a) Magnetic field at the bottom of the trap $B_0$~[G] as a function of the dimensionless $IB_{var}$ parameter (see text). 
(b) to (i) Measurement of the Larmor frequency. Atom numbers in state $\left|b\right>$ and $\left|2,2\right>$, $N_{\left|b\right>}+N_{\left|2,2\right>}$, in arbitrary unit as a function of the frequency shining the atomic cloud $f_{knife}$~[MHz] for several values of the $IB_{var}$ parameter, the red open circles are experimental data and the solid blue line are fits (see text).}
\end{figure*}

Figure \ref{fig_Trap_Bottom} gathers all the measurements done for the calibration of the magnetic of the traps used in section \ref{sec_ExpChooseMag}.

\begin{figure*}
\centering  \includegraphics[width=1.00\textwidth]{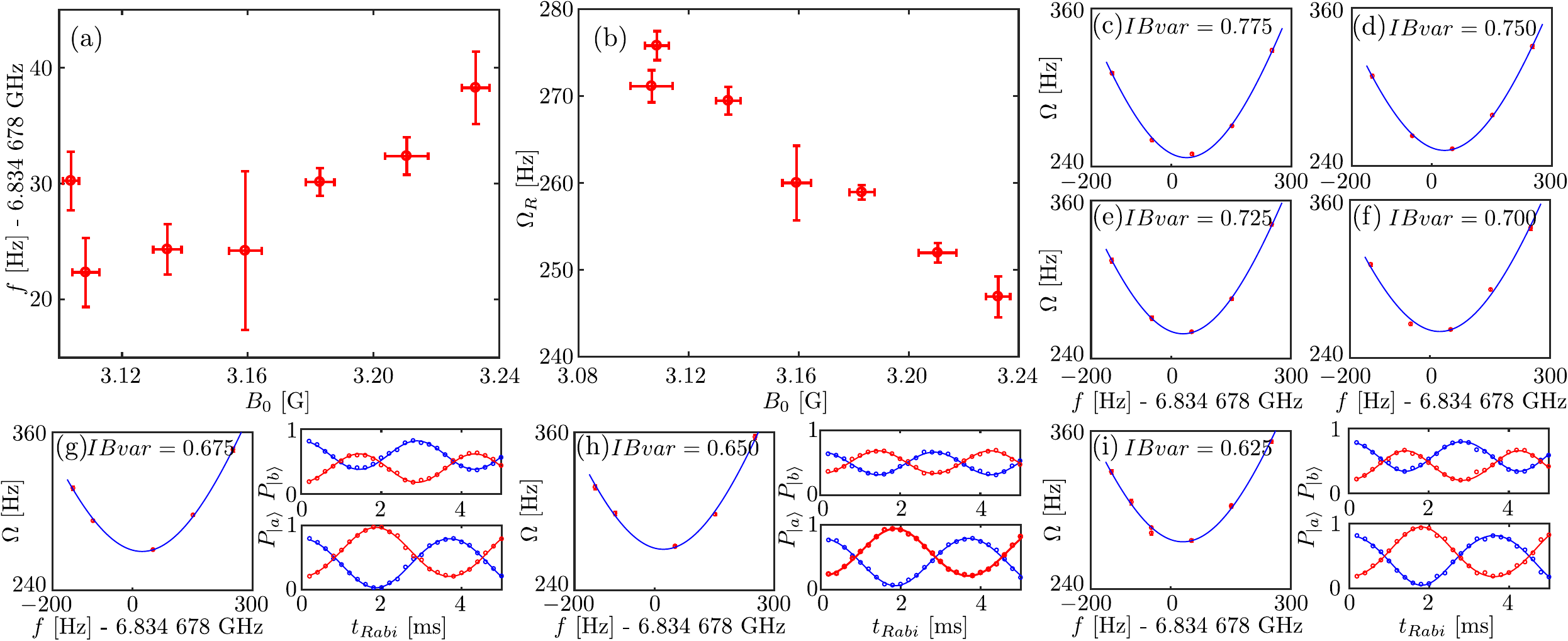}
\caption{\label{fig_Rabi_Osc} (Color online) Determination of the Rabi oscillations parameters - frequency of the two photon field and pulse duration - for the traps of figure \ref{fig_Trap_Bottom}. 
(a) Frequency of the local oscillator $f=f_{RF}+f_{MW}$~[Hz] used for driving Rabi oscillations at their minimum oscillation frequency $\Omega_R$~[Hz] as function of the magnetic field at the bottom of the trap $B_0$~[G]. 
(b) $\Omega_R$~[Hz] as a function of $B_0$~[G]. 
(c) to (i) Rabi oscillation frequency $\Omega$~[Hz] as a function of the frequency of the local oscillator driving the Rabi oscillations $f$~[Hz] for several values of the $IB_{var}$ parameter (see text), the open circles are experimental data and the solid blue lines are fits (see text). 
Inset in (g), (h) and (i) show examples of Rabi oscillations as a function of the Rabi time $t_{Rabi}$~[ms]. 
Population of state $\left|b\right>$ (respectively $\left|a\right>$) is in blue (respectively in red), the open circles are experimental data and the solid lines are fits (see text).
Although the data (a) seems to show a linear relation between $A_1$ and $B_0$ it is important to note that this is not true \cite{Treutlein2008}.}
\end{figure*}

Figure \ref{fig_Rabi_Osc} shows the experimental finding of the Rabi pulse parameters for the traps of figure \ref{fig_Trap_Bottom}.

\begin{figure*}
\centering  \includegraphics[width=1.00\textwidth]{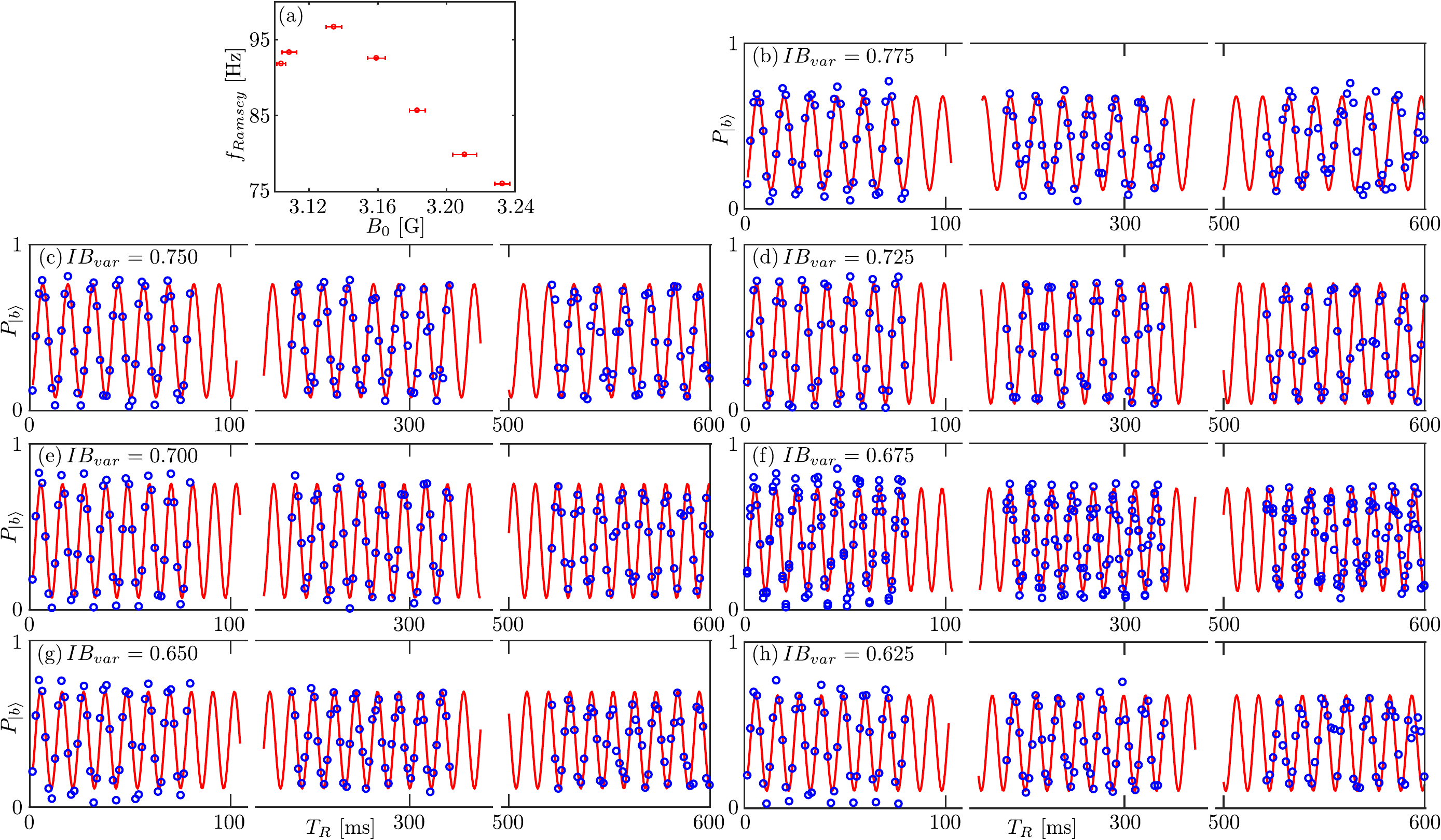}
\caption{\label{fig_RamseyFringes} (Color online) Ramsey fringes recorded in the traps of figure \ref{fig_Trap_Bottom} with the Rabi pulse parameters found in figure \ref{fig_Rabi_Osc}. 
(a) Frequency of the Ramsey fringes $f_{Ramsey}$~[Hz] as a function of the magnetic field at the bottom of the trap $B_0$~[Hz]. 
(b) to (h) Ramsey fringes, population of state $\left|b\right>$, $P_{\left|b\right>}$, as a function of the Ramsey time $T_R$~[ms] for several values of the $IB_{var}$ parameter (see text). 
The blue open circles are the experimental data and the red solid lines are fits (see text).}
\end{figure*}

Figure \ref{fig_RamseyFringes} shows all the Ramsey fringes recorded in the traps of figure \ref{fig_Trap_Bottom}.

%\newpage
\bibliography{biblio}

\end{document}